\documentclass[a4paper,preprintnumbers,floatfix,superscriptaddress,pra,twocolumn,showpacs,notitlepage,longbibliography
]{revtex4-2}
\usepackage{bbm}
\usepackage[T1]{fontenc}
\usepackage{inputenc}
\usepackage{color}
\usepackage{babel}
\usepackage{physics}
\usepackage{mathtools}
\usepackage{bbding}
\usepackage{pifont}
\usepackage{textcomp}
\usepackage{wasysym}
\usepackage{amsthm}
\usepackage[normalem]{ulem}
\usepackage{nicefrac}
\usepackage{wasysym}
\usepackage{dsfont}
\usepackage{amsmath}
\usepackage{amssymb}
\usepackage{graphicx}
\usepackage{hyperref}
\usepackage{cleveref}
\usepackage{amsfonts}
\usepackage{enumerate}
\usepackage{orcidlink}
\usepackage{txfonts}
\hypersetup{
     colorlinks   = true,
     linkcolor    = blue,
     citecolor    = blue,
     urlcolor     = blue,
     }
\DeclareSymbolFont{ptmxcalletters}{OMS}{ztmcm}{m}{n}
\DeclareMathAlphabet{\mathcal}{OMS}{ztmcm}{m}{n}

\newtheoremstyle{mystyle}
  {0pt}      
  {0pt}      
  {} 
  {\parindent}         
  {\itshape} 
  {.}        
  {.5em}     
  {\thmname{#1}\thmnumber{ #2}\thmnote{ {\itshape(#3)}}}

\theoremstyle{mystyle}
\newtheorem{theorem}{Theorem}

\newtheorem{lemma}{Lemma}

\newtheorem{corollary}{Corollary}

\newtheorem{definition}{Definition}

\begin{document}
\title{Higher-order quantum information inequalities and error-correction of non-Markovian noise}

\date{\today}
\author{Matheus Capela\,\orcidlink{0000-0002-9207-1177}}

\email{capela.quantum@gmail.com}
\affiliation{Department of Physics and Astronomy, University of Rochester, Rochester, New York 14627, USA}
\affiliation{Departamento de F\'isica, Universidade Federal de S\~{a}o Carlos, Rodovia Washington Lu\'is, km 235 - SP-310, 13565-905 S\~{a}o Carlos, SP, Brazil}

\date{\today}
\begin{abstract}
Defining entropy of a quantum channel has become a challenge with potential candidates recently reported in literature. In this Article, it is taken under consideration the entropy of a quantum channel defined as the quantum entropy of its associated Choi state. The consequences presented here of such approach is two-fold. First, the limitations on the information-processing of quantum channels are assessed, leading to the \emph{higher-order quantum information inequalities}. Those are then considered as useful witnesses of non-Markovian effects beyond a given number of quantum side-channels in a quantum process. Importantly, the presentation is given with the application of the methods for the study of examples. Second, such approach provides a characterization on the condition for the recovery of quantum channel preparations under non-Markovian noise. Such result is a straightforward extension of standard error-correction conditions which gives further meaning to the higher-order information measures defined here. 
\end{abstract}
\maketitle

\section{Introduction}
The understanding of the limitations imposed on communication tasks is ubiquitous in information theory, independently of the underlying theory being classical or quantum  \cite{wilde2013quantum}. In this regards, a groundbreaking concept concept is that of the entropy, characterizing the optimal compression a source can undergo without changing its defining features \cite{shannon1948mathematical,schumacher1995quantum}.

A recent trend in the formulation of quantum theory is the understanding of quantum channels taking place as fundamental building blocks \cite{taranto2025higher,costa2026indefinite}. That is motivated by other elementary operations being simply instances of quantum channels with particular forms for the input and output systems \cite{chiribella2009theoretical}: A quantum state preparation is a quantum channel with trivial (one-dimensional) input system; a quantum measurement is a quantum channel with classical output system; any temporal evolution is represented by some appropriate quantum channel. 

As a consequence, it became relevant the extension of concepts related to quantum states to quantum channels. That goes from the uncertainty principle \cite{xiao2021uncertainty} to quantum entanglement \cite{gour2021entanglement,theurer2312single} to quantum thermodynamics \cite{badhani2510thermodynamic,badhani2026conditional,badhani2026erasure} and beyond \cite{chen2020entanglement,jencova2021general,wang2019quantifying,yokojima2021consequences}. Not surprisingly, the definition of entropy for channels holds as a relevant problem. In this regard, it stands out the approach of Ref.~\cite{gour2021entropy}, in which the authors define the entropy of a quantum channel from a relative entropy of quantum channels previously defined in \cite{cooney2016strong,leditzky2018approaches}. Interestingly, such approach gives as a consequence a quantum channel merging theorem \cite{gour2021entropy}. 

In turn, the authors in Ref.~\cite{kurzyk2021relating} study how defining the entropy of a quantum channel as the entropy of its Choi state representation relates with the alternative in \cite{gour2021entropy}. There, their main result is to show that the former implies a upper-bound on the later.  
This Article goes in the direction of further understanding the information-theoretic properties for which defining the entropy of a quantum channel as the entropy of its correspondent Choi state would imply. The result  of such approach is two-fold, as described in the following. 

First, we show that such definition for the entropy of quantum channels naturally leads to the higher-order version of quantum information inequalities. In turn, such information-theoretic bounds on quantum information processing impose limitations on quantum processes with a restricted side-channel structure. Precisely, it is determined the interventional strategy for which probing a local quantum system accordingly gives further information on the nature of its relying environment. Importantly, the methods developed are addressed in practice in the study of toy models.

Second, standard error-correction conditions previously reported in Ref.~\cite{schumacher1996quantum} can be reformulated in terms of the higher-order formalism developed here, in such a way that it shows how time-correlated noisy quantum gate preparations could be error-corrected. This is a topic with some recent interest due to the limitations on near-term quantum devices \cite{ziyad2025emergent,f2025detrimental,kobayashi2024tensor}. Interestingly, the approach considered here is complementary to recent similar results in literature \cite{sohail2026fundamental,tanggara2024strategic}.

The Article is organized as follows. Section~\ref{section_2} provides a presentation on the literature necessary for the understanding of the results. That makes the presentation nearly self-contained. Section~\ref{section_3} presents a discussion on the definition entropy of a quantum channel addressed here. Special attention is given to how it related with previous attempts in literature. Section \ref{section_4} presents the main result: The higher-order quantum information inequalities. Section \ref{section_5} shows how the adopted entropy provides a suitable characterization of non-Markovian noise error-correction. Finally, Section~\ref{section_conclusion} presents the final remarks on the presented results.

\section{Quantum processes}\label{section_2}

In this section we present the notation, definitions and results related to the process formalism necessary in the Article. To every quantum system $A$ there is an associated Hilbert space, which we denote here with the same symbol. An isomorphic copy of a quantum system $A$ is denoted by $\overline{A}$. The quantum systems under consideration in this study are finite-dimensional, and the dimension of a quantum system $A$ is denoted as $d_{A}$.

The space of linear transformations with input vectors in $A$ and output vectors in $B$ is denoted with $L(A;B)$. In case the input and output are isomorphic, we simply write $L(A)$. The identity operator acting on $A$ is denoted by $\mathds{1}_{A}$. Importantly, the state $\rho$ of a quantum system $A$ is a positive semi-definite operator in $L(A)$ with unit trace. In turn, The collection of linear maps $\Lambda:L(A) \rightarrow L(B)$ is denoted as $T(A;B)$. The identity map acting on the system $A$ is denoted with $\mathrm{id}_{A}$. A quantum channel is a completely positive and trace-preserving (CPTP) map in $T(A;B)$. The interested reader is urged to follow the standard Ref.~\cite{wilde2013quantum} for a detailed discussion on the elementary concepts of quantum theory.

Importantly, a linear supermap is a trasformation of the form $\Upsilon: T(A;B) \rightarrow T(P;F)$. The identity supermap is denoted as $\mathrm{id}_{A \rightarrow B}$. Here, the set of all linear supermaps is denoted by $P(A,B;C,D)$. A (single-slot) quantum process is a linear supermap $\Upsilon \in P(A,B;C,D)$ for which the following conditions hold \cite{chiribella2008transforming}:
\begin{enumerate}[(i)]
\item If $\Lambda \in T(A \otimes C; B \otimes D)$ is a completely positive map, then $(\Upsilon \otimes \mathrm{id}_{C \rightarrow D})(\Lambda)$ is a completely positive map. 
\item If  $\Lambda \in T(A;B)$ is trace preserving, then $\Upsilon(\Lambda)$ is trace preserving.

\end{enumerate}

Single-slot quantum processes have a circuit realization. Precisely, if $\Upsilon$ is a quantum process in $P(A,B;C,D)$, then there is an isometric quantum channel $\mathcal{V}$ in $T(C;A \otimes E)$ and a unitary quantum channel $\mathcal{U}$ in 
$T(B \otimes E; D \otimes E^{\prime})$ for which holds \cite{chiribella2008transforming}
\begin{equation}
\Upsilon(\Lambda)(\cdot) = \Tr_{E'}[ \mathcal{U} \circ (\Lambda \otimes \mathrm{id}_{E})\circ \mathcal{V}(\cdot)],
\end{equation}
for every input $\Lambda$. Hence, there is a non-Markovian model for every single-slot quantum process. 

The space of multilinear supermaps of the form $\Upsilon: T(A_1;B_1) \times \cdots \times T(A_n;B_n)  \rightarrow T(C;D)$ is denoted with $P(A_1,B_1:A_2,B_2:\dots:A_n,B_n;C,D)$. To each multilinear supermap  $\Upsilon \in P(A_1,B_1:A_2,B_2:\dots:A_n,B_n;C,D)$ there is associated a unique linear supermap $\Upsilon^{\prime} \in P(A_1\otimes \cdots \otimes A_n, B_1 \otimes \cdots B_n;C,D)$ such that $\Upsilon(\Lambda_1,\dots,\Lambda_n)=\Upsilon^{\prime}(\Lambda_1\otimes \cdots\otimes \Lambda_n)$, for any input linear maps $\Lambda_k \in T(A_k;B_k)$ with $k=1,\dots,n$ \cite{greub1967multilinear}.
Multi-slot quantum processes are multilinear supermaps $\Upsilon$ with associated linear supermap $\Upsilon^{\prime}$ such that
\begin{equation}
\begin{split}
   \left( \Upsilon 
   \otimes  \left( \mathrm{id}_{C_1\rightarrow D_1} 
   \otimes \cdots \otimes 
   \mathrm{id}_{C_n\rightarrow D_n}
   \right) 
   \right)(\Lambda_1\otimes \cdots \otimes \Lambda_k) \\ \text{is a quantum channel},
\end{split}
\end{equation}
for each quantum channel $\Lambda_k \in T(A_k\otimes C_k; B_k\otimes D_k)$ with $k\in \{1,\dots,n\}$ \cite{oreshkov2012quantum,araujo2015witnessing}. Note that the condition above is a simple extension of the definition of single-slot quantum  processes to the multipartite case.  

It is worth to mention that, in distinction with the single-slot case, multi-slot quantum processes do not have a non-Markovian implementation in general \cite{oreshkov2012quantum}. In this regard, quantum processes with fixed causal order, also called by quantum combs for the form of their pictorial representations, constitute a strict subset of the space of quantum processes. Importantly, this Article aims to address such a class of processes, for which side-channels playing the role of quantum memories arise naturally.

Finally, let us define the Choi state associated with a quantum channel $\Lambda: L(A) \rightarrow L(B)$ as the bipartite quantum state in $L(B \otimes A)$ as \cite{wilde2013quantum}
\begin{equation}\label{eq:Choi_repr}
    \lambda=(\Lambda \otimes \textrm{id}_{A})(\phi^{+}),
\end{equation}
where $\phi_{A\overline{A}}^{+}$ denotes the maximally entangled state. This channel-state corresponce plays crucial role in the results reported in this Article. Namely, it allows the concepts related to quantum states to be extended to their higher-order form, and vice versa. In the following Sections we show that this naive approach gives non-trivial bounds on quantum information-processing. 

\section{Entropy of a quantum channel}\label{section_3}

Here, we aim to discuss on a suitable definition for the entropy of a quantum channel, that is, the transformation of quantum systems. As mentioned above, every elementary quantum operation, such as quantum state preparations and quantum measurement devices, can be understood as a particular form of a quantum channel. That motivates the investigation for the entropy of a quantum channel.

Importantly, it is well-known that every quantum channel $\Lambda_{A\rightarrow B}$ is equivalently represented by (isomorphic to) its associated Choi $\lambda_{BA}$ state. In fact, the correspondence in Eq.~(\ref{eq:Choi_repr}) is a valid representation of linear maps as linear operators. Following this reasoning, we consider the entropy of a channel as equal to the entropy of its associated Choi state. The precise definition adopted here is given as follows.

\begin{definition}[Entropy of a quantum channel]
Let $\Lambda$ denote the quantum channel with input and output systems $A$ and
$B$, respectively. Then, we denote its entropy with $H(A\rightarrow B)_{\Lambda}$, and make the assignment
\begin{equation} \label{eq:def_ent}
    H(A\rightarrow B)_{\Lambda} \coloneqq H(A\otimes B)_{\lambda},
\end{equation}
where $\lambda$ is the Choi state associated to $\Lambda$. The entropy of a quantum system $X$ prepared in the state $\rho$ is defined as $H(X)=-\Tr[\rho \log_2 \rho]$.
\end{definition}

It is worth to mention that several distinct definitions for the entropy of a quantum channel have been considered in the literature (see Ref.~\cite{Roga2011} and references therein). For instance, Ref.~\cite{gour2021entropy} propose a definition for the entropy of a quantum channel based upon a quantum relative entropy for channels. Importantly, their definition satisfy the assumptions previously reported in Ref.~\cite{gour2019comparison}, and it determines the capacity for the quantum channel merging protocol. 

In turn, the definition considered here has been previously compared with the one in Ref.~\cite{gour2021entropy}, and reported in Ref.~\cite{kurzyk2021relating}. There, the quantity in Eq.~(\ref{eq:def_ent}) is proven to be an upper-bound on the entropy of Ref.~\cite{gour2021entropy}. Moreover, this bound is also known to be saturated for  unital qubit quantum channels ~\cite{kurzyk2021relating}. Nevertheless, no further information-theoretic result is discussed there.

In the present Article we provide such desirable properties of a genuine quantum information measure. Importantly,
we show in the next sections that the definition adopted here imply several desired results. Firstly, it naturally induces a correspondence between quantum information inequalities and their higher-order version (that is, information inequalities for the transformation of channels instead of states). Secondly, it sets an appropriate framework for the statement of a necessary and sufficient condition for error-correction of time-correlated noise. 

\section{Information theory in the process formalism}\label{section_4}

\subsection{Brief review on classical and quantum information inequalities} \label{inf_ineqs_section}
Markovianity is ubiquitous feature of (stochastic) communication processes in classical information theory \cite{shannon1948mathematical}, as it allows for the mathematical description of time-local operations. For the sake of presenting a precise definition, let us consider a classical process, consisting of a sequence of random variables $X_1,X_2,X_3$ drawn with a joint probability distribution. The process is Markovian whenever the variable $X_3$ is independent of the variable $X_1$ when given $X_2$ \cite{yeung2008information}. That is, the Markov condition states that a complete prediction of the future depends solely on the knowledge of the present, and thence, is independent of the past.

The Markov condition can clearly be extended to processes with arbitrary number of random variables in straightforward manner. Let $X_1,\dots,X_n$ be drawn from a probability distribution $p$, for some natural number $n \geq 3$. The process is Markovian, and denoted with $X_1 \rightarrow X_2 \rightarrow \cdots \rightarrow X_n$, if it holds the condition \cite{yeung2008information}
\begin{equation}
p(X_k|X_1,\dots,X_{k-1})=p(X_k|X_{k-1}),
\end{equation}
for all $k \geq 3$. 

In turn, information inequalities arise in the context of classical information theory as limitations on the admissible correlations spread in time by classical Markovian processes.  For instance, any Markovian process $X_1 \rightarrow X_2 \rightarrow X_3$ satisfy the (classical) information inequality \cite{yeung2008information}
\begin{equation}\label{dpi_1}
I(X_1;X_2) \geq I(X_1;X_3),
\end{equation}
where the correlations between random variables is measured with their mutual information defined as $I(X_i,X_j)\coloneqq H(X_i)+H(X_j)-H(X_i,X_j)$, with respect to the Shannon entropy $H(X)_p \coloneqq -\sum_{x:p(x) \neq 0}p(x)\log_2 p(x)$. 

In its classical domain,  Markovianity is preserved under time reversal \cite{yeung2008information}. That is, every forward Markovian process $X_1 \rightarrow X_2 \rightarrow \cdots \rightarrow X_n$ implies that its correspondent backwards process $X_1 \leftarrow X_2 \leftarrow \cdots \leftarrow X_n$ also forms a Markov process. Then, for a three-steps forward Markovian process $X_1 \rightarrow X_2 \rightarrow X_3$, we must also have the validity of the information inequality \cite{yeung2008information}
\begin{equation}\label{dpi_2}
I(X_2;X_3) \geq I(X_1;X_3).
\end{equation} 

Considering the form of the information inequalities in Eqs.~(\ref{dpi_1},\ref{dpi_2}) reveals that for any $n$-steps Markovian process $X_1 \rightarrow \cdots \rightarrow X_n$ it follows the data processing inequality \cite{yeung2008information}
\begin{equation}
I(X_i;X_j) \geq I(X_r;X_s),
\end{equation}
for $\{i,\dots,j\} \subset \{r,\dots,s\}$, with $i<j$ and $r<s$. It is worth mentioning that such condition is instrumental in the proof of the fundamental theorem in classical Shannon theory (see Sect. 7 in Ref.~\cite{yeung2008information}). 

Similarly, quantum information inequalities are limitations on quantum Markov processes with respect to information measures \cite{wilde2013quantum}. For instance, let us consider as figure of merit the coherent information, defined as \cite{schumacher1996quantum}
\begin{equation}
I_c(\rho;\Lambda) \coloneqq H(B)_{\Lambda(\rho)}-H(B\otimes R)_{(\Lambda\otimes\mathrm{id}_{R})(\psi)},
\end{equation}
for the state preparation $\rho$ of a quantum system $A$ with purification $\psi$ with respect to an extended system $A \otimes R$, and a quantum channel $\Lambda:L(A) \rightarrow L(B)$. 

Let us consider the three-step quantum Markov process $\rho_1 \rightarrow \rho_2\rightarrow \rho_3$, with  $\rho_2 \coloneqq \Lambda_1(\rho_1)$ and $\rho_3 \coloneqq \Lambda_2(\rho_2)$ for fixed quantum channels $\Lambda_1,\Lambda_2$ (with arbitrary input and output systems). Then, it holds the quantum data-processing inequality \cite{schumacher1996quantum}
\begin{equation}\label{qdpi}
    I_c(\rho_1;\Lambda_1) \geq I_c(\rho_1;\Lambda_2 \circ \Lambda_1),
\end{equation}
for any initial quantum state preparation $\rho_1$.

The quantum information inequality in Eq.~(\ref{qdpi}) is a quantum analogue of Eq.~(\ref{dpi_1}). In general, classical information inequalities serve as patterns for the development of quantum information-processing conditions. That is the main reasoning under the original formulation in Ref.~\cite{schumacher1996quantum}. Interestingly, the backwards information inequality in Eq.~(\ref{dpi_2}) has no quantum version in general. That is, its correspondent inequality, $I_c(\Lambda_1(\rho_1);\Lambda_2) \geq I_c(\rho_1;\Lambda_2 \circ \Lambda_1)$, holds in general only for isometric quantum channels $\Lambda_1$ \cite{hayashi2017quantum}.

\subsection{Higher-order quantum information inequalities}
Now, we show that higher-order quantum operations inherits several properties with respect to the entropy of quantum channels defined in Eq.~(\ref{eq:def_ent}). In particular, the Schumacher-Nielsen quantum information inequalities \cite{capela2022quantum} (as much as the quantum error-correction condition from Ref.~\cite{schumacher1996quantum}) take the higher-order form. Moreover, it reveals useful tools for the operational characterization of non-Markovian quantum processes. 

\begin{figure}[t!] 
\centering   
\includegraphics[width=0.48\textwidth]{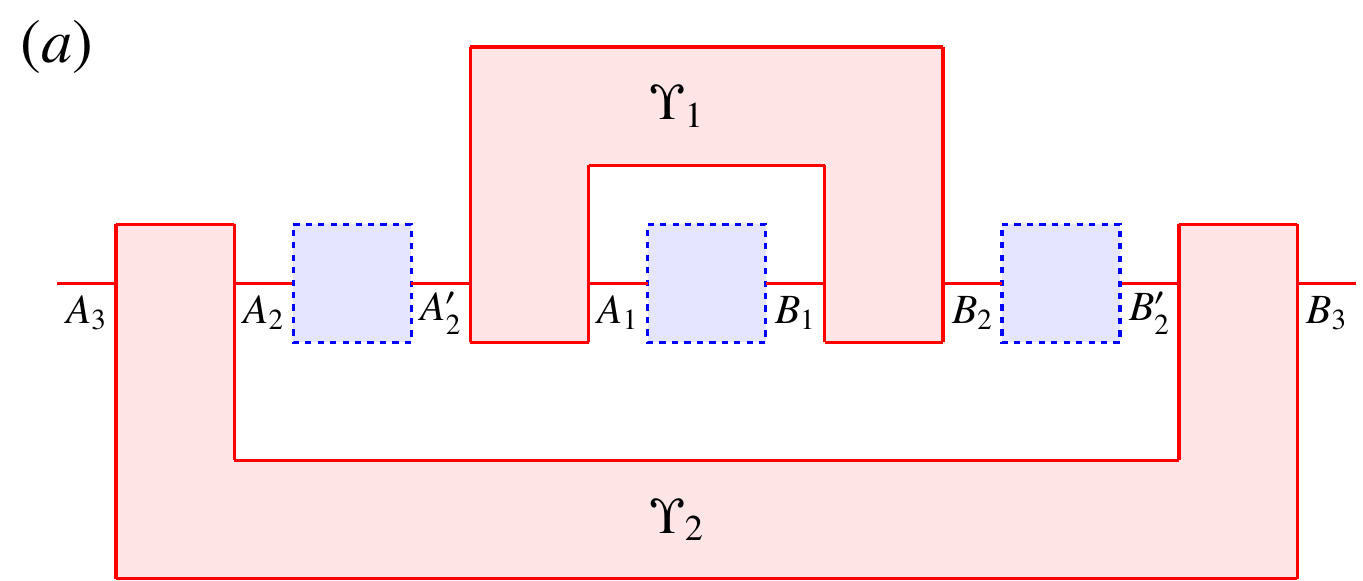} 
\includegraphics[width=0.48\textwidth]{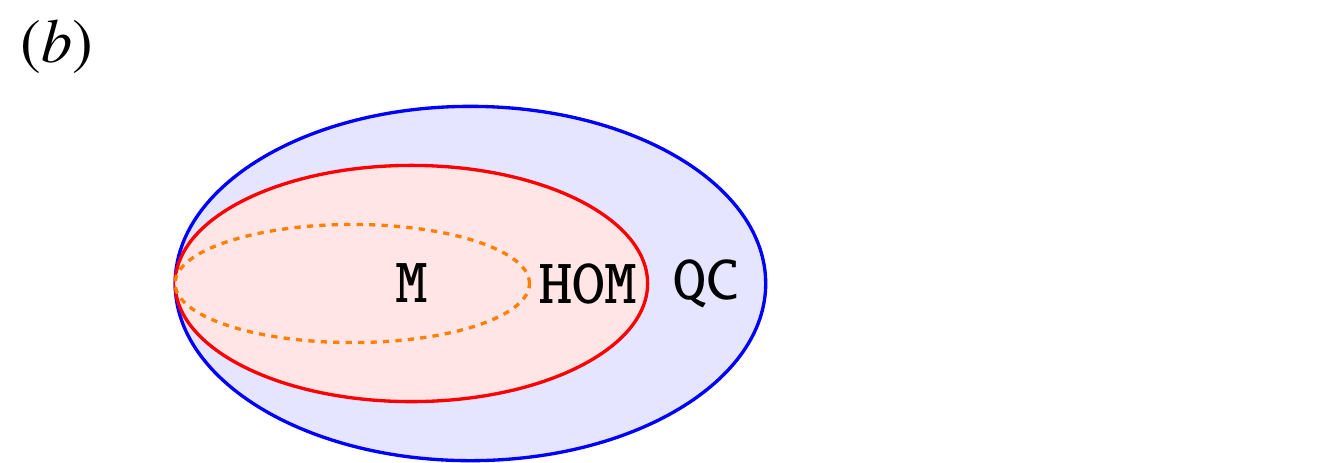} 
\caption{Higher-order quantum Markov condition. (a) The diagram provides a pictorial representation of a three-slots quantum comb for which the higher-order Markov condition holds. Moreover, it follows the validity of the higher-order quantum information inequality in Eq.~(\ref{channel-processing_ineq}) for every input quantum channel from $A_1$ to $B_1$. The dashed regions in shaded blue color represent the open slots for interventions, allowing for the assessment of the process under information extracted from the local system only. (b) The bottom panel is a Vein diagram representing the (set) inclusion relations between the processes considered here. Memoryless quantum processes, also called Markovian and denoted with $\mathtt{M}$ in the diagram (region delimitated with the orange dashed curve), are higher-order Markovian (denoted as $\mathtt{HOM}$) with trivial environmental systems. Importantly, a violation of the higher-order information inequalities witness that the quantum process lies outside the red region in the diagram. It is worth to mention that neither $\mathtt{M}$ nor $\mathtt{HOM}$ are convex sets. The set of quantum combs, that is, quantum multitime processes are denoted with $\mathtt{QC}$.}
    \label{figure1}
\end{figure}

For the following discussion, let us consider the sequence of transformations of quantum channels $\Lambda_1 \rightarrow \Lambda_2 \rightarrow \cdots \rightarrow \Lambda_n$. The $k$-th quantum channel $\Lambda_k:L(A_k) \rightarrow L(B_k)$ is mapped into $\Lambda_{k+1}:L(A_{k+1})\rightarrow L(B_{k+1})$ by a quantum superchannel denoted with $\Upsilon_k$. This is a higher-order analogue of the quantum Markov condition. Nonetheless, it allows for limited memory effects (see Fig.~\ref{figure1}). Interestingly, such form of quantum processes, defined from the successive action of quantum superchannels, also emerges in foundational considerations in quantum theory \cite{lie2023quantum}. Here, their information-theoretic properties are addressed as follows.

It is convenient to study higher-order operations from the perspective of Choi operators \cite{oreshkov2012quantum}. Here, we denote the Choi quantum state of each channel $\Lambda_k$ with $\lambda_k$. The transformation $\Upsilon_k: \Lambda_k \mapsto \Lambda_{k+1}$ induces a completely positive and trace-preserving quantum channel $\Delta_k: \lambda_k \mapsto \lambda_{k+1}$ \cite{chiribella2008transforming}. Thus, the quantum process $\lambda_1 \rightarrow \lambda_2 \rightarrow \cdots \rightarrow \lambda_n$ reflects in several information-theoretic limitations on the higher-order quantum process $\Lambda_1 \rightarrow \Lambda_2 \rightarrow \cdots \rightarrow \Lambda_n$. We take into scrutiny such conclusions in the following.

Here, we aim to investigate the consequences of similar reasoning on higher-order quantum operations. In particular, it can be defined an extension of the concept of quantum coherent information, now, for a quantum channel $\Lambda$ with respect to a single-slot process $\Upsilon$,
\begin{multline} 
    \widetilde{I_c}(\Lambda;\Upsilon)=H(A\rightarrow B)_{\Upsilon(\Lambda)}-H(A \rightarrow B\otimes R)_{(\Upsilon \otimes \mathrm{id}_{R})(\mathcal{V})},
\end{multline}
where $\mathcal{V}_{A\rightarrow B\otimes R}$ is an isometric dilation of $\Lambda$, that is, an isometric quantum channel for which holds $\Lambda(\cdot)=\Tr_R [\mathcal{V}(\cdot)]$.

Importantly, it follows -- from the Choi representation of the higher-order process -- that it holds a higher-order quantum information inequality.
\begin{theorem}[Higher-order quantum information inequality]
Let $\Upsilon_1$ and $\Upsilon_2$ be quantum processes in $P(A_1,B_1;A_2,B_2)$ and $P(A_2,B_2;A_3,B_3)$, respectively. Then, for every quantum channel $\Lambda$ in $T(A_1;B_1)$ it holds that
\begin{equation}\label{channel-processing_ineq}
    \widetilde{I_c}(\Lambda;\Upsilon_1)\geq \widetilde{I_c}(\Lambda;\Upsilon_2 \circ \Upsilon_1).
\end{equation}
\end{theorem}

The interested reader is urged into considering Appendix~\ref{appendix_proof} for a detailed presentation of the proof of Eq.~(\ref{channel-processing_ineq}). The following discussion focus on its consequences and applications.

In a sense, the condition above would allow us to detect environmental side-channels in well-defined steps of the quantum process. Namely, let us consider a 3-slot quantum process $\Upsilon$ as defined in Fig.~\ref{figure1}. Then, we could check its non-Markovianity with the violation of ordinary quantum data-processing inequalities \cite{schumacher1996quantum}. Interestingly, with the higher-order information inequality in Eq.~(\ref{channel-processing_ineq}) we could witness the non-divisibility $\Upsilon\neq \Upsilon_2\circ \Upsilon_1$ \footnote{\label{fn:divisibility}To be precise, the correct condition would be the product form of the process matrix as $W=W_{A_1A_2^{\prime}B_1B_2}^{(1)}\otimes W_{A_2A_3B_2^{\prime}B_3}^{(2)}$. The reader is referred to Ref.~\cite{oreshkov2012quantum} for further information on the process matrix formalism.}. Moreover, in that case, the non-Markovianity is surely arising from specific environmental side-channels from $A_3$ to $B_3$ and from $A_2$ to $B_2$ (see Fig.~\ref{figure1}(a)). Thus, a violation of the higher-order quantum information inequality witness the effects of side-channels beyond those accounted in the higher-order Markov model (see Fig.~\ref{figure1}(b)). 

Since higher-order quantum information inequalities follow directly from the ordinary Markovianity of the Choi representation of the process, the quantum Markov monogamy inequalities, complementary quantum information bounds reported in Ref.~\citep{capela2022quantum}, also take place in their higher-order form. Nevertheless, as that is not crucial for the understanding of the methods developed here, those are presented in Appendix~\ref{appendix_Markov_monogamy}. Now, it is taken under scrutiny how higher-order quantum information inequalities provide useful characterization of non-Markovian quantum processes.

\section{Witnessing quantum non-Markovian effects beyond higher-order quantum Markovianity}\label{witness} 
The higher-order quantum information inequality provides a practical operational method to witness non-Markovian phenomena beyond what is predicted by the higher-order quantum Markov model. That is, there is a particular scheme for detecting such non-Markovian effects by probing a local system in open quantum dynamics \cite{milz2021quantum}. In the following, this interventional approach is considered \footnote{The terminology `interventional approach` refers to acquiring information from a quantum process by acting locally on a target subsystem with any quantum operation defined from a quantum instrument. That is, in such approach one neglects any information relying upon the knowledge of environmental states, which is assumed to be experimentally inaccessible. See Ref.~\cite{milz2021quantum} for a more broad discussion on such approach in quantum information science.}. 

\begin{figure}[t!] 
\centering   
\includegraphics[width=0.48\textwidth]{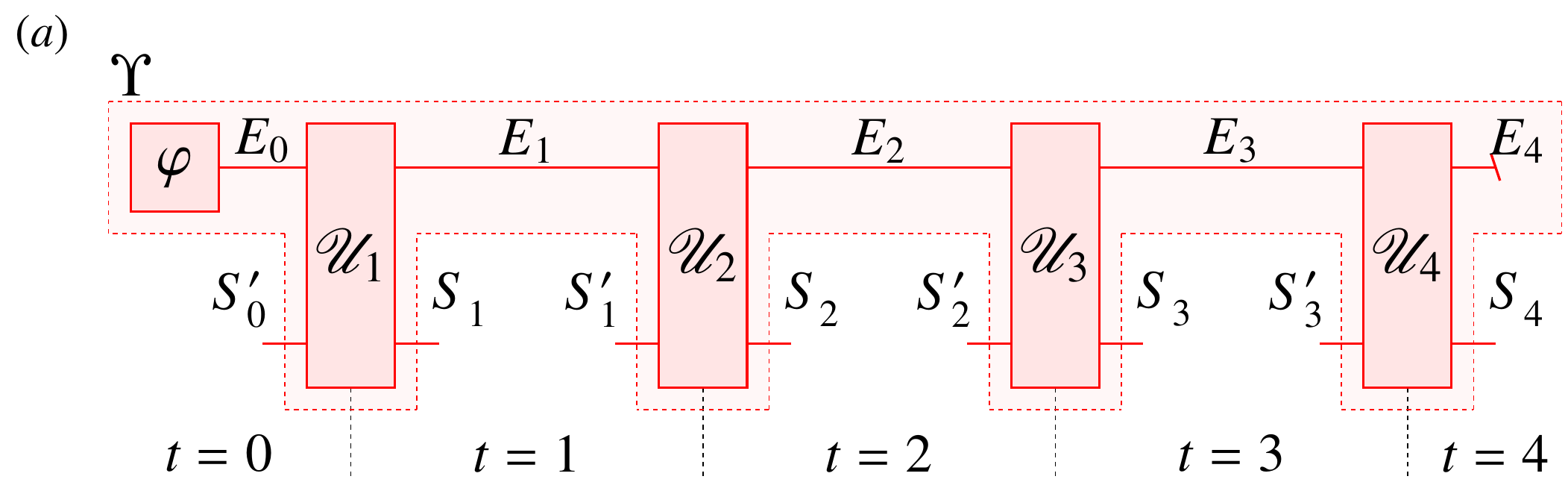} 
\includegraphics[width=0.48\textwidth]{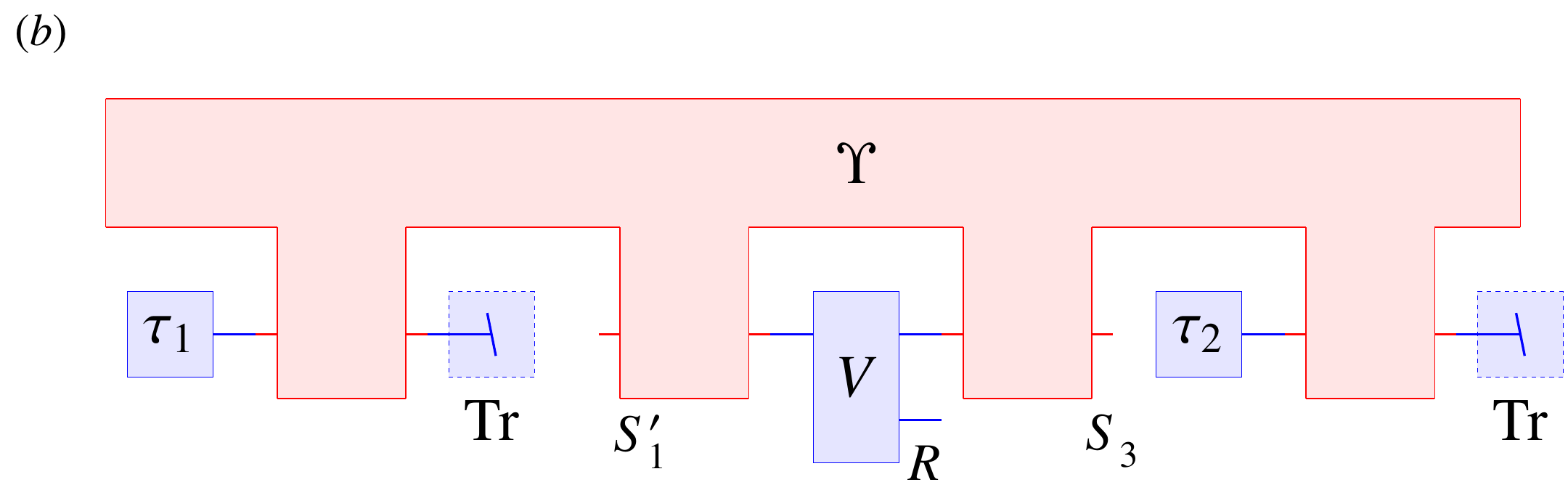} 
\includegraphics[width=0.48\textwidth]{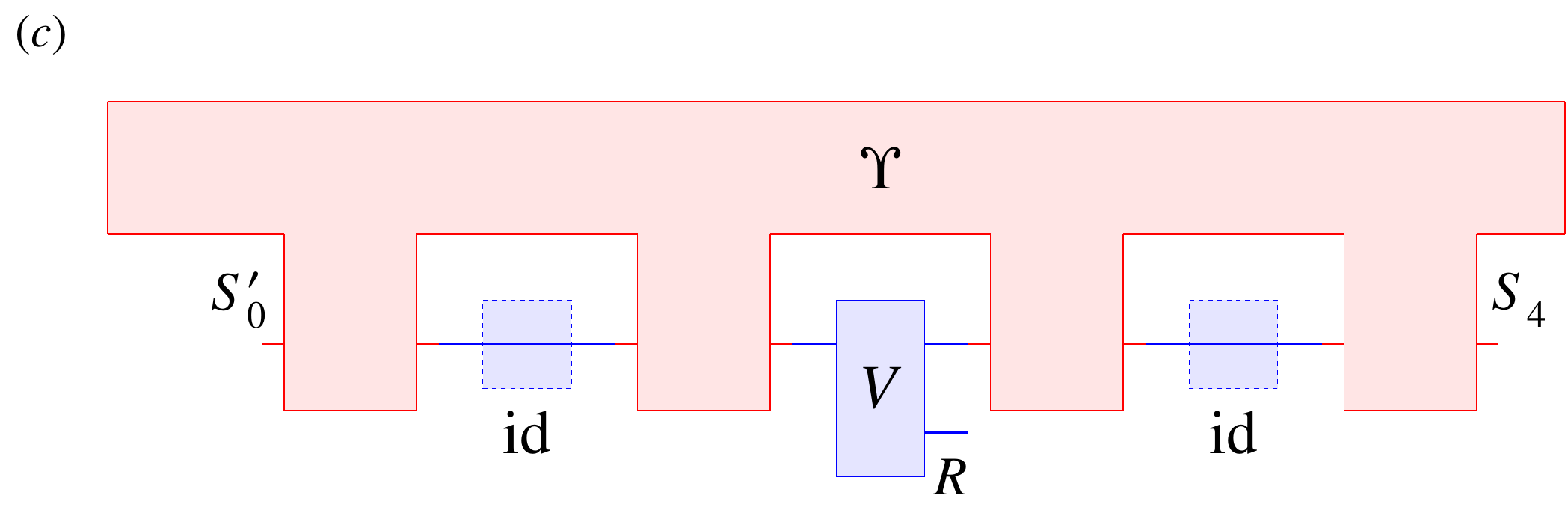} 

\caption{Three-slot non-Markovian quantum processes and quantum interventions $\mathcal{I}_1$ and $\mathcal{I}_2$. Panel~(a) shows a diagrammatic representation of an arbitrary five-step non-Markovian quantum process $\Upsilon$, generated with a sequence of unitary system-environment channels. In the main text, the example addressed in detail is defined with each unitary transformation $\mathcal{U}_{k}$ represented in Eq.~(\ref{unitary_matrix}). The environment is initially prepared on the pure quantum state $\varphi=\ket{+}\bra{+}$. Panels~(b) and (c) show the diagrams defining the intervention strategies $\mathcal{I}_1$ and $\mathcal{I}_2$, respectively. The particular intervention addressed in the main text is defined with $\tau_1=\tau_2=\ket{0}\bra{0}$, and $\mathcal{V}(\bullet)=\textsf{swap}(\bullet \otimes \omega)\textsf{swap}^{\dagger}$. }
    \label{fig:Sec_V_1}
\end{figure}

Let us examine in detail three-slot quantum processes, as defined in the diagram presented in Fig.~\ref{fig:Sec_V_1}(a). Such quantum combs are implemented with system-environment quantum dynamics allowing for local interventions on the system-part. Hence, the quantum processes $\Upsilon$ under scrutiny can be understood as a five-step process with time instants denoted with $0 \leq t \leq 4$. Moreover, we denote the input and output space of the intervention on the system-part at each instant of time $t$ with $S_{t}^{}$ and $S_{t}^{\prime}$, respectively.

Here, we consider the particular example of a global two-qubit environment-system unitary quantum channel $\mathcal{U}(\cdot)=U(\cdot)U^{\dagger}$, where the single-Kraus operator is represented in the standard basis $\{\ket{0},\ket{1}\}$ by
\begin{equation}\label{unitary_matrix}
U=\begin{bmatrix}
0 & -\sqrt{1 - \eta} & \sqrt{\eta} & 0 \\
1 & 0 & 0 & 0 \\
0 & 0 & 0 & 1 \\
0 & \sqrt{\eta} & \sqrt{1 - \eta} & 0
\end{bmatrix},
\end{equation}
with $0 \leq \eta \leq 1$. The dynamics considered is homogeneous in the sense that the global time evolution is given by the same unitary quantum channel for every time-step. The environment starts in the pure quantum state preparation $\ket{\varphi}=(\ket{0}+\ket{1})/\sqrt{2}$. See Fig.~\ref{fig:Sec_V_1}(a) for a pictorial representation.

Importantly, our goal is to implement an intervention on the process in such a way that we could be sure of its incompatibility with higher-order quantum Markovianity by the violation of Eq.~(\ref{channel-processing_ineq}). The desired protocol is defined in two steps as follows.

Firstly, we choose any isometric quantum channel $\mathcal{V}:L(S_2^{}) \rightarrow L(S_2^{\prime}\otimes R)$, which plays the role of the input quantum channel in the higher-order quantum information inequality when discarding the auxiliary system $R$, that is, $\Lambda=\Tr_R \circ \mathcal{V}$. Then, we take quantum state preparations $\tau_1$ and $\tau_2$ for $S_0^{\prime}$ and $S_3^{\prime}$, respectively. Feeding each quantum system of $\Upsilon$ with the correspondent quantum state or operation defined above, and discarding the quantum systems $S_1^{}$ and $S_4^{}$, it gives us a quantum channel from $S_1^{\prime}$ to $S_3^{} \otimes R$. Such quantum channel, denoted by $\Omega_1$, is represented with the diagram in Fig.~\ref{fig:Sec_V_1}(b), and plays the role of $\Upsilon_1(\Lambda)$ in the term $\widetilde{I_c}(\Lambda;\Upsilon_1)$ of Eq.~(\ref{channel-processing_ineq}) for a higher-order Markov model. The intervention defined in this first step is denoted as $\mathcal{I}_1$.

Secondly, we take the same isometric quantum channel $\mathcal{V}$ previously chosen in  $\mathcal{I}_1$. Then, we feed the slots in steps $t=1$ and $t=3$ with identity quantum channels. The resulting quantum channel, denoted by $\Omega_2$, is represented in Fig.~\ref{fig:Sec_V_1}(c), and plays the role of $\Upsilon_2 \circ \Upsilon_1(\Lambda)$ in the term $\widetilde{I_c}(\Lambda;\Upsilon_2 \circ \Upsilon_1)$ for a higher-order Markov model. This second intervention is denoted with $\mathcal{I}_2$.

The validity of the higher-order quantum information inequality can be equivalently stated by the following mathematical assertion.

\begin{corollary}[Interventional form of the higher-order quantum information inequality] \label{thm_1}
Let $\Upsilon$ be a five-steps quantum process (three-slots quantum comb) for which the higher-order Markov condition holds (Fig.~\ref{figure1}(a)). Then, let $\Omega_1$ and $\Omega_2$ be the quantum channels generated with the interventions $\mathcal{I}_1$ and $\mathcal{I}_2$ defined in Fig.~\ref{fig:Sec_V_1}, respectively. It follows that
\begin{multline}\label{eq:thm_1}
H(S_1^{\prime} \rightarrow S_3^{})_{\Tr_{R} \circ \, \Omega_1}-H(S_1^{\prime} \rightarrow S_3^{} \otimes R)_{\Omega_1} \\
\geq H(S_0^{\prime} \rightarrow S_4^{})_{\Tr_R \circ \, \Omega_2}-H(S_0^{\prime} \rightarrow S_4^{} \otimes R)_{\Omega_2}.
\end{multline} 
\end{corollary}

Finally, the abovementioned procedure suffices to verify the higher-order non-Markovianity of $\Upsilon$. In order to check the violation of Eq.~(\ref{eq:thm_1}) it is necessary to compute the entropy of quantum channels, which effectively reduces to the entropy of their Choi states. The detailed discussion on the computation of the entropy of the relevant quantum channels $\Omega_1,\Omega_2$ are presented in Appendix~\ref{appendix_intervention}. 

Now, we assess how the higher-order quantum information inequality taking place in its interventional formulation, presented in Corollary~\ref{thm_1}, performs on witnessing non-Markovian effects in the particular model considered here. 

\begin{figure}[t!] 
\centering   
\includegraphics[width=0.48\textwidth]{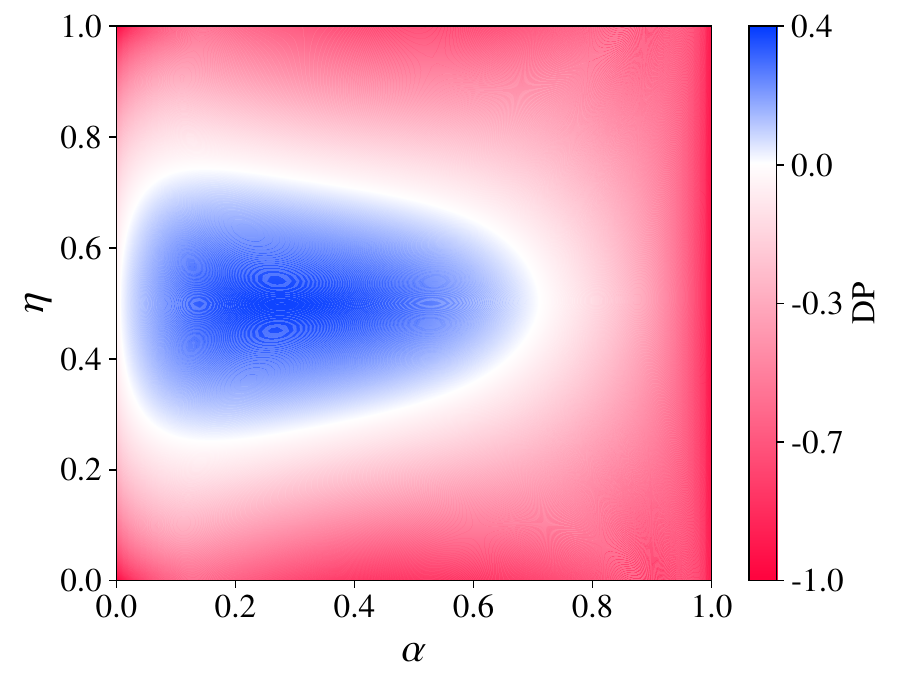} 

\caption{Violation of the higher-order quantum information inequality in Eq.~(\ref{eq:thm_1}) for the quantum process defined in Fig.~\ref{fig:Sec_V_1}(a) with unitary layers as in Eq.~(\ref{unitary_matrix}) and initial environmental state preparation $\ket{+}$. The plot presents the quantity $\mathrm{DP}$ as a function of the interventional quantum state preparation parameter $\alpha$ and the process parameter $\eta$. The interventional quantum states $\tau_1$ and $\tau_2$ are fixed as $\ket{0}$. Interestingly, the higher-order quantum information inequality is violated for a wide range of the parameters $\mu$ and $\alpha$, witnessing that the quantum process under scrutiny has quantum side-channels beyond the expected by the higher-order Markov model.}
    \label{fig:plot_1}
\end{figure}

Figure~\ref{fig:plot_1} shows how much Eq.~({\ref{eq:thm_1}}) is violated with negative values, quantity denoted as $\textrm{DP}$. The adopted isometric intervention is defined as $\mathcal{V}(\cdot)=\textsf{swap}(\cdot \otimes \omega)\textsf{swap}^{\dagger}$, with the swap bipartite quantum gate defined from the relation $\textsf{swap}_{AB}\ket{i}_A \otimes \ket{j}_B = \ket{j}_B \otimes \ket{i}_A$, where $\ket{i}$ with $i\in \{0,\dots,
d_A-1\}$, and $\ket{j}$ with $j\in \{0,\dots,d_B-1\}$, constitute orthonormal basis of $A$ and $B$, respectively. Furthermore, the intervention $\mathcal{V}$ keeps the input quantum state of its correspondent slot while feeding its output with the quantum state $\omega$. Here, it is considered a pure quantum state preparation $\omega=\ket{\omega}\bra{\omega}$, with $\ket{\omega}=\sqrt{1-\alpha}\ket{0}+\sqrt{\alpha}\ket{1}$ depending upon a real-valued parameter $0 \leq \alpha \leq 1$. Importantly, the interventional quantum state preparations $\tau_1=\tau_2=\ket{0}\bra{0}$ are also fixed.

Finally, it is worth to note that the parameter $\eta$ is fixed for a given process. On the other hand, the parameter $\alpha$ defining the interventions $\mathcal{I}_1$ and $\mathcal{I}_2$ is completely arbitrary. For the model considered here, there are values of $\eta$ for which some range of $\alpha$ is not able to detect the higher-order non-Markovianity of the process. Nevertheless, for each fixed $\eta$, there is a value of $\alpha$ such that Eq.~(\ref{eq:thm_1}) is violated. Furthermore, there is always an intervention scheme for which one would be successful in the desired task for the model under question.

\subsection{An alternative form of higher-order quantum Markovianity}The method of witnessing environmental effects on the probed quantum system developed above is robust enough to allow us to consider a distinct type of higher-order quantum Markov condition. Here, the important condition is for a process to have a limited number of quantum side-channels. A complete discussion of this topic is left to Appendix~\ref{alternative_HOQMC}.



\subsection{Higher-order quantum information inequalities versus ordinary quantum information inequalities}The study and development of quantum information-processing limitations is a longstanding problem in quantum information theory with recent developments. For instance, in Ref.~\cite{capela2022quantum} the Author along with collaborators were able to show that there are quantum information inequalities apart from the quantum data-processing inequality \cite{schumacher1996quantum}. Interestingly, those are not equivalent information principles, and it is possible to find a violation of the former while the later is still valid for particular quantum non-Markovian processes. That shows that the further development of quantum information inequalities might provide useful characterization of non-Markovian phenomena.

Hence, one may argue that it would be relevant to assess how the novel higher-order quantum information inequalities outperform the previously mentioned methods on witnessing non-Markovianity. In that sense, it is important to note that such comparison is not entirely reasonable in the sense that violations of the distinct type of inequalities provide a characterization of different regions in the space of quantum processes. That is, a violation of the ordinary quantum information inequalities witness process beyond Markovianity, while a violation of higher-order quantum information inequalities witness processes beyond the higher-order Markovianity. See Fig.~\ref{figure1}(b) for a graphical schematics.

\subsection{Higher-order quantum Markov monogamy inequality}As mentioned above, the higher-order quantum information inequality is not the only quantum information inequality one should expect to hold. In Appendix~\ref{appendix_Markov_monogamy}, it is presented another non-trivial higher-order quantum information inequality.

\section{Error-correction of a quantum channel under non-Markovian noise}\label{section_5}

Finally, it is worth to highlight an application of the quantum error-correction theorem to be first reported in Ref.~\cite{schumacher1996quantum} to the context of higher-order quantum operations: The higher-order quantum information inequality reveals the regime for which non-Markovian noise can be error-corrected. In the following discussion it is made precise such a claim.

For completeness, let us briefly review the Nielsen-Schumacher error-correction result. Suppose we prepare the state $\rho$ of a quantum system $A$ as the input of a quantum channel $\Lambda_{A\rightarrow B}$. Then, under what condition could we have an error-correction channel at the output of such transformation? The main result in Ref.~\cite{schumacher1996quantum} provides an answer to this question. There, it was shown that there is a quantum channel $\mathcal{R}_{B \rightarrow \overline{A}}$ for which $\mathcal{R} \circ \Lambda$ acts as the identity channel on every pure state $\psi$ on the support of $\rho$. 

Now, we aim to discuss a higher-order formulation of the result in Ref.~\cite{schumacher1996quantum}. In principle, the simple reformulation of their theorem in terms of higher-order quantum operations (the process formalism) would provide an interesting result, namely, the condition under which non-Markovian noise acting on the implementation of unitary gates could be error-corrected. This would be of relevance in face of the current noisy performance of quantum computational devices \cite{ziyad2025emergent,f2025detrimental,kobayashi2024tensor}. 

\begin{figure}
\centering   
\includegraphics[width=0.4\textwidth]{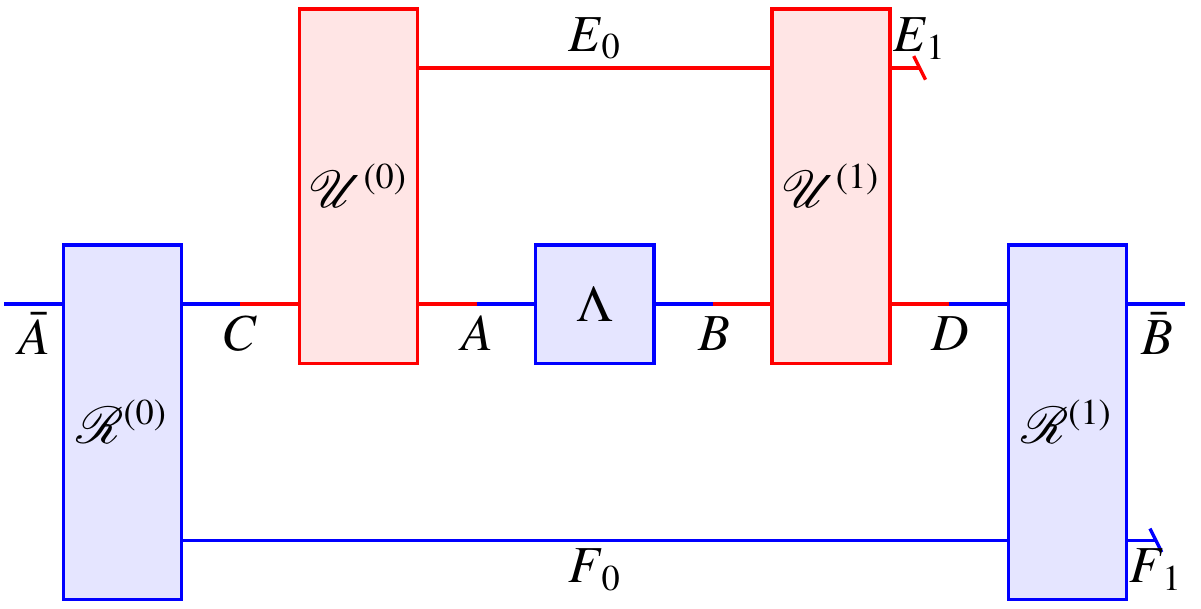} 
\caption{Non-Markovian error effects on the implementation of a quantum channel. In the diagram, the target quantum channel $\Lambda$ is affected by the influence of time-correlated quantum noise $\Upsilon$, implemented with the isometric channel $\mathcal{U}_{C\rightarrow AE_0}^{(0)}$ and unitary channel $\mathcal{U}_{E_0B\rightarrow E_1D}^{(1)}$. In turn, the error-correction is performed by the quantum superchannel $\widetilde{\mathcal{R}}$, implemented with isometric and unitary channels $\mathcal{R}_{A \rightarrow C F_0}^{(0)}$ and $\mathcal{R}_{D F_0\rightarrow BF_1}^{(1)}$, respectively.}
    \label{figure2}
\end{figure}

Suppose we aim to implement a quantum channel $\Lambda$. Nevertheless, due to some imperfection on the preparation of such operation, we are given with some distinct channel $\Upsilon(\Lambda)$ for some quantum comb $\Upsilon$. In face of the realization theorem of quantum superchannels, the noisy version of our target quantum operation can be understood as the result of some non-Markovian noise. How could we error-correct such time-correlated noise? It is possible to make use of the Nielsen-Schumacher error-correction theorem described above straightforwardly: There is an error-correction quantum superchannel $\widetilde{\mathcal{R}}$ for which $\widetilde{\mathcal{R}} \circ \Upsilon$ acts trivially on each quantum isometric gate $\mathcal{U}_{A\rightarrow B}$ in the Choi support of $\Lambda$ if and only if (see Fig.~\ref{figure2}) 
\begin{equation}
H(A \rightarrow B)_{\Lambda} = I_c(\Lambda ; \Upsilon),
\end{equation}  
where we say that a quantum channel $\mathcal{E}^{(1)}$ is in the Choi support of $\mathcal{E}^{(2)}$ whenever the Choi state of the former is in the support of the Choi state of the later.

The proof for such a statement is given as follows. Let $\lambda_{AB}$ denote the Choi state of the channel $\Lambda_{A \rightarrow B}$. The quantum superchannel $\Upsilon$ induces some quantum channel representation on the level of Choi operators which we denote here as $\Delta_{AB\rightarrow CD}$ (see discussion in Appendix~\ref{appendix_proof} and Refs.~\cite{chiribella2008transforming,gour2019comparison}). Then the Nielsen-Schumacher error-correction theorem determines that there is a quantum channel $\mathcal{R}_{CD \rightarrow \overline{AB}}$ such that $(\mathcal{R} \circ \Delta)(\varphi)=\varphi$ for each pure quantum state $\varphi_{AB}$ in the support of $\lambda$. Finally, note that a pure Choi state $\varphi_{AB}$ is associated with some isometric quantum channel $\mathcal{U}_{A \rightarrow B}$. That completes the proof of the claim.

\section{Conclusion}\label{section_conclusion}

Here, the consequences of adopting the entropy of a Choi state as an entropy measure of its associated quantum channel are taken under scrutiny. Hence, the results reported can be understood as complementary to the ones reported in Ref.~\cite{gour2021entropy} in which the authors consider a distinct definition for the entropy of a quantum channel based on a relative entropy of quantum channels. 

Great attention is given to the higher-order quantum information inequalities, that is, the information-theoretic limitations on the transformation of quantum channels. The approach adopted here is similar to standard considerations in quantum information theory for which information inequalities motivates the definition of quantum information measures \cite{schumacher1996quantum}.  

The results are also used in the certification of non-Markovian effects beyond a determined fixed number of side channels, enforcing constraints on quantum memories in a process. It is worth to mention that such approach provides better understanding on the role of multitime interventions on quantum information inequalities. In this sense, the present work goes forward in the open problems reported in Ref.~\cite{capela2022quantum}. There, quantum information inequalities following from two-step interventions are developed for the certification of arbitrary non-Markovian effects. Here, it is shown that three-step interventions are able to provide further detailed information on the environmental structure in a quantum process. Precisely, a violation of the higher-order quantum information inequalities would detect quantum side-channels not present in a higher-order Markov model. Nevertheless, the role of multitime interventions in general is still an open problem to be addressed. 

Interestingly, the approach considered in this Article leads to a better understanding of the conditions under which non-Markovian quantum noise can be error-corrected: Time-correlated noise on the implementation of any isometric quantum channel in the Choi support of a reference quantum channel can be error-corrected whenever its entropy is equal to the coherent information with respect to the noisy operation. This consideration is complementary to recent results reported in Refs.~\cite{sohail2026fundamental,tanggara2024strategic}. It would be desirable to further characterize the implementation of the error-correction of time-correlated noise under physically motivated sources of error, nevertheless, this is left for future work.

\acknowledgements
\noindent
I thank Kaumudibikash Goswami for helpful discussions.
This study was financed, in part, by the São Paulo Research
Foundation (FAPESP), Brasil. Process Numbers 2024/06972-9, 2022/00209-6 and 2025/15372-8.

\appendix

\section{Proof of the higher-order quantum information inequality}\label{appendix_proof}
Here, we provide a formal derivation of Eq.~(\ref{channel-processing_ineq}).
\begin{proof}
Let $\Lambda_1 \rightarrow \Lambda_2 \rightarrow \Lambda_3$ be a higher-order quantum Markov process. Equivalently, there is an input quantum channel $\Lambda_1 \in T(A_1;B_1)$ and single-slot quantum combs $\Upsilon_1 \in P(A_1,B_1;A_2,B_2),\Upsilon_2 \in P(A_2,B_2;A_3,B_3)$ for which holds $\Lambda_2=\Upsilon_1(\Lambda_1)$ and $\Lambda_3=\Upsilon_2(\Lambda_2)$.

Now, let us denote the Choi state of $\Lambda_k$ as $\lambda_k$ with $k=1,2,3$. Then, there is a completely positive and trace-preserving linear map $\Delta_1 : L(B_1 \otimes A_1) \rightarrow L(B_2 \otimes A_2)$ such that $\lambda_2 = \Delta_1(\lambda_1)$ \cite{chiribella2008transforming,gour2019comparison}. Similarly, there is a quantum channel $\Delta_2 : L(B_2 \otimes A_2) \rightarrow L(B_3 \otimes A_3)$ for which $\lambda_3 = \Delta_2(\lambda_2)$. Thus, $\lambda_1 \rightarrow \lambda_2 \rightarrow \lambda_3$ forms a quantum Markov process. Furthermore, it holds the quantum data-processing inequality \cite{schumacher1996quantum}
\begin{equation}
I_c(\lambda_1;\Delta_1) \geq I_c(\lambda_1;\Delta_2 \circ \Delta_1). 
\end{equation}

The inequality above is precisely the desired result. Note that a possible purification $\psi_1 \in L(B_1 \otimes R \otimes A_1)$ consists in the Choi state of the isometric extension $\mathcal{V}: L(A_1) \rightarrow L(B_1 \otimes R)$ of $\Lambda_1$, that is,
\begin{equation}
\psi_1 = (\mathcal{V} \otimes \mathrm{id}_{A_1})(\phi^{+}).
\end{equation}

Then, it holds that 
\begin{widetext}
\begin{eqnarray}
H(B_2 \otimes A_2 )_{\Delta_1(\lambda_1)} = H(A_2 \rightarrow B_2)_{\Upsilon_1(\Lambda_1)},\\ 
H(B_3 \otimes A_3)_{\Delta_2 \circ \Delta_1(\lambda_1)} = H(A_3 \rightarrow B_3)_{\Upsilon_2 \circ \Upsilon_1(\Lambda_1)},\\ 
H(B_2 \otimes R \otimes A_2)_{(\Delta_1 \otimes \mathrm{id}_{R})(\psi_1)} = H(A_2 \rightarrow B_2 \otimes R)_{(\Upsilon_1 \otimes \mathrm{id}_{R})(\mathcal{V})},\\
H(B_3 \otimes R \otimes A_3)_{((\Delta_2 \circ \Delta_1 )\otimes \mathrm{id}_{R})(\psi_1)} = H(A_3 \rightarrow B_3 \otimes R)_{((\Upsilon_2 \circ \Upsilon_1 )\otimes \mathrm{id}_{R})(\mathcal{V})}.
\end{eqnarray}
\end{widetext}
Note that the quantities $H(B_j \otimes A_j)-H(B_j \otimes R \otimes A_j)$, with $j=2,3$, are independent on the choice of purification $\psi_1$ \cite{schumacher1996quantum}.

\end{proof}

\section{Computation of the Choi states associated with the quantum channels in Fig.~\ref{fig:Sec_V_1}.} \label{appendix_intervention} 
Now, we consider how to compute the entropy of the quantum channels $\Omega_1$ and $\Omega_2$ in Figs.~\ref{fig:Sec_V_1}(a) and \ref{fig:Sec_V_1}(b), respectively. That is, we propose a circuit implementation of the correspondent Choi states from which the quantum entropy of channels can be directly computed. In the following presentation, it is worth to consider that for every isometric quantum channel $\mathcal{V}:L(A)\rightarrow L(B \otimes R^{\prime})$ there is some unitary quantum channel $\mathcal{V}':L(A \otimes R^{})\rightarrow L(B \otimes R^{\prime})$ and a unique pure quantum state preparation $\omega$ for which holds $\mathcal{V}(\cdot)=\mathcal{V}'(\cdot \otimes \omega)$. 

\begin{figure}[t] 
\centering   
\includegraphics[width=0.5\textwidth]{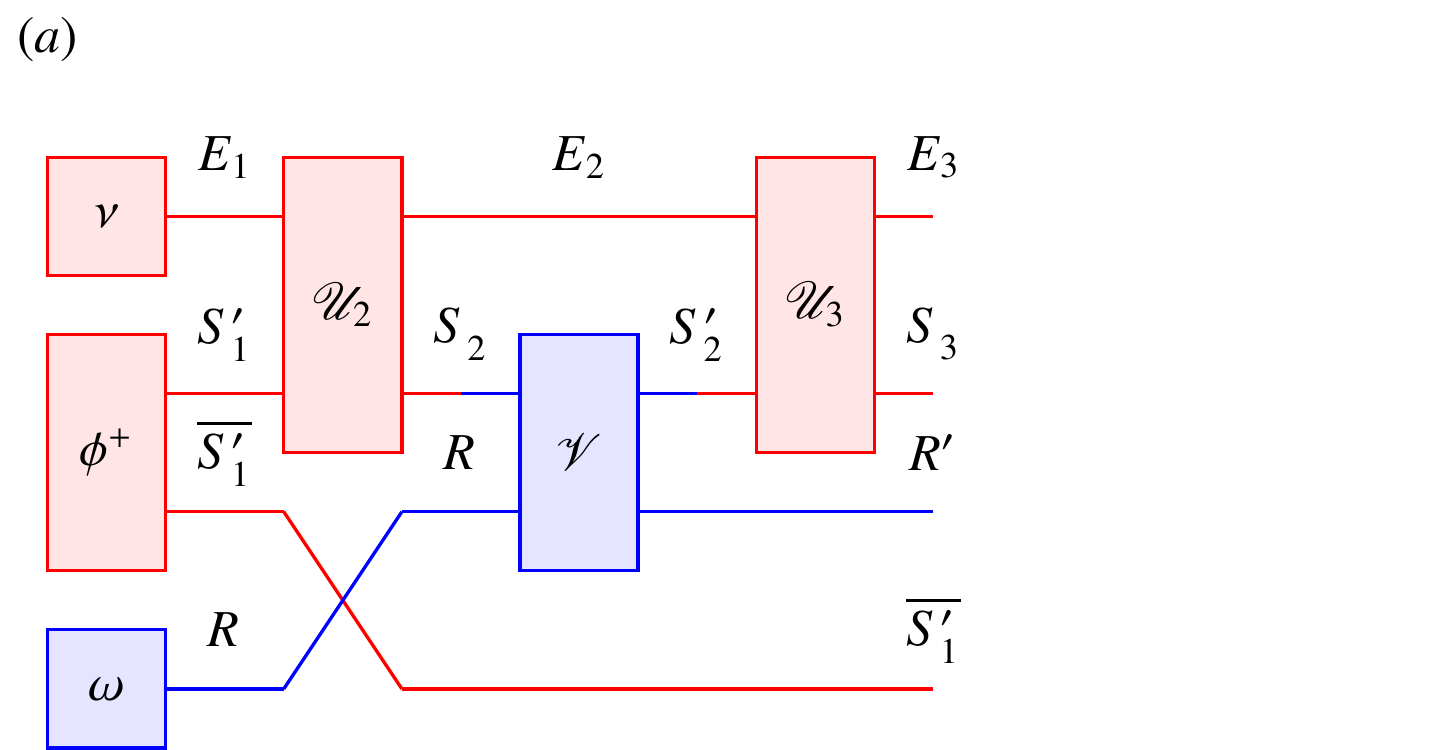} 
\includegraphics[width=0.5\textwidth]{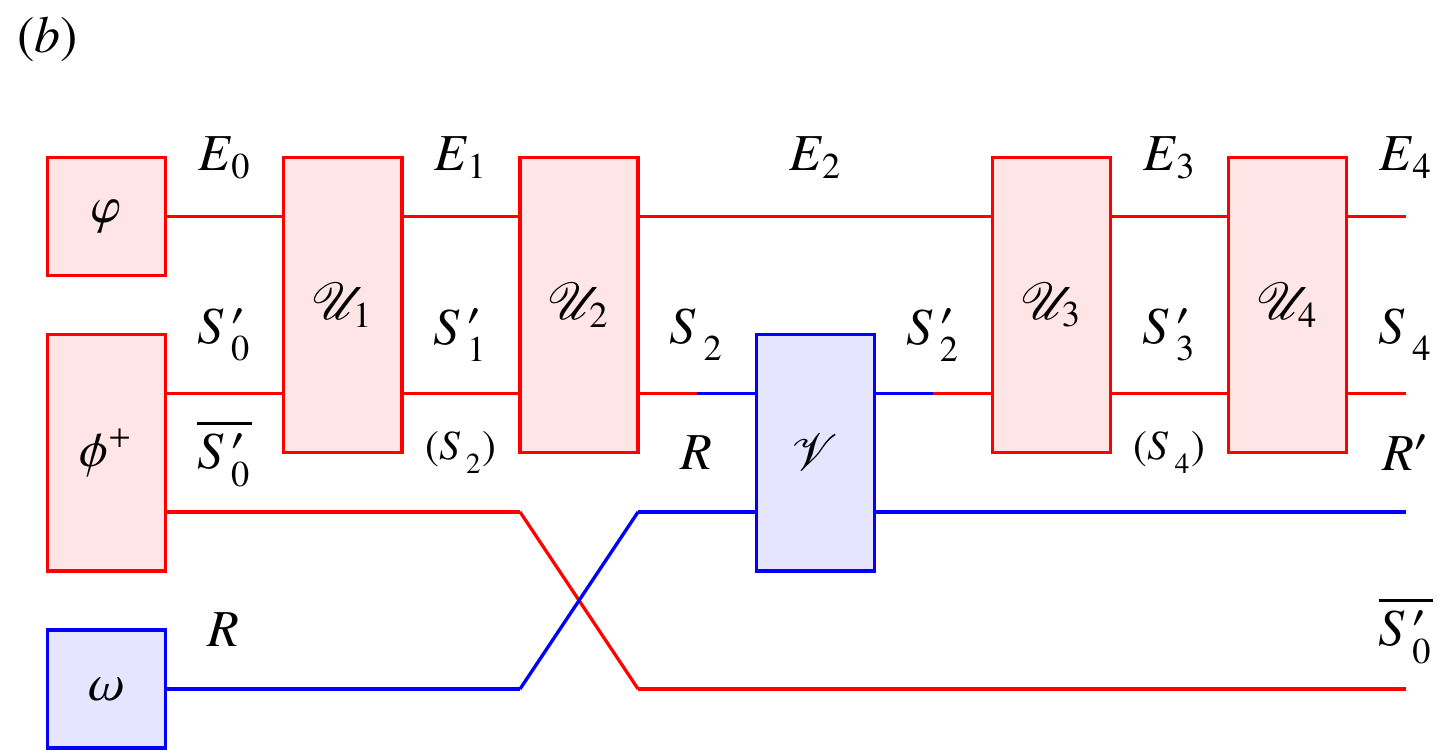} 
\caption{Choi states of the quantum channels $\Omega_1$ and $\Omega_2$. Figures (a) and (b) show the diagram representation of the Choi states of the quantum channels generated with the intervention $\mathcal{I}_1$ and $\mathcal{I}_2$, respectively.}
    \label{fig:choi_states}
\end{figure}

Then, let us start considering the quantum channel $\Omega_1$. Its Choi state can be computed with the following sequence of quantum operations, as illustrated in Fig.~\ref{fig:choi_states}(a). 
\begin{enumerate}
\item Prepare the environment system $E_1$ in the quantum state $\nu=\Tr_{S_1^{}}[\mathcal{U}_1(\varphi \otimes \tau_1)]$. Then, prepare the four-partite quantum state
\begin{equation}
\rho_1= \nu \otimes \phi^{+} \otimes \omega,
\end{equation}
where $\phi^{+}$ is the maximally entangled state of $S_1^{\prime} \otimes \overline{S_1^{\prime}}$.
\item Take the local action of the unitary quantum channel $\mathcal{U}_2$,
\begin{equation}
\rho_2=(\mathcal{U}_2 \otimes \mathrm{id}_{\overline{S_1^{\prime}} \otimes R^{}})(\rho_1).
\end{equation}
\item Change the order of the local systems $\overline{S_1^{\prime}}$ and $R^{}$,
\begin{equation}
\rho_3=(\mathrm{id}_{E_2 \otimes S_2^{}} \otimes \mathsf{SWAP}_{\overline{S_1^{\prime}} R^{}})(\rho_2),
\end{equation}
where the swap quantum channel is defined as $\mathsf{SWAP}(\cdot
)=\mathsf{swap}(\cdot)\mathsf{swap}^{\dagger}$.
\item Act locally with the unitary quantum channel $\mathcal{V}^{\prime}$ on the correspondent subsystems,
\begin{equation}
\rho_4 = (\mathrm{id}_{E_2} \otimes \mathcal{V}^{\prime} \otimes \mathrm{id}_{\overline{S_1^{\prime}}})(\rho_3).
\end{equation}
\item Take the local transformation $\mathcal{U}_3$ with
\begin{equation} \label{eq:Choi_Omega_1_E}
\rho_5 =  (\mathcal{U}_3 \otimes \mathrm{id}_{R^{\prime} \otimes \overline{S_1^{\prime}}})(\rho_1).
\end{equation}
\item Finally, discarding the environment system of the quantum state in Eq.~(\ref{eq:Choi_Omega_1_E}) gives the Choi state of the desired quantum channel,
\begin{equation} \label{eq:Choi_Omega_1}
\rho_6=\Tr_{E_3}[\rho_5].
\end{equation}
\end{enumerate}

Note that in the presentation of Thm.~\ref{thm_1} it was not necessary to explicitly consider the input reference system $R^{}$, and hence, the superscript of the output reference system $R^{\prime}$ was dropped. The quantum state in Eq.~(\ref{eq:Choi_Omega_1}) is the Choi state of the desired quantum channel $\Omega_1$. Furthermore, it is possible to compute the entropic terms in the left-hand-side (l.h.s.) of Eq.~(\ref{eq:thm_1}) from $\rho_6$.

Now, let us define the protocol for computing the Choi state of $\Omega_2$. The reader is urged to follow the protocol with the assistance of the diagram in Fig.~\ref{fig:choi_states}(b).

\begin{enumerate}
\item Prepare the environment system $E_0$ in the state $\varphi$. Then, prepare the four-partite quantum state
\begin{equation}
\sigma_1= \varphi \otimes \phi^{+} \otimes \omega,
\end{equation}
where $\phi^{+}$ denotes the maximally entangled state of $S_0^{\prime} \otimes \overline{S_0^{\prime}}$. 
\item Take the local action of the first two unitary channels,
\begin{equation}
\sigma_2 = ((\mathcal{U}_2 \circ \mathcal{U}_1) \otimes \mathrm{id}_{\overline{S_0^{\prime}} \otimes R^{}} )(\rho_1).
\end{equation}
\item Change the order of the subsystems $\overline{S_0^{\prime}}$ and $R^{}$,
\begin{equation}
\sigma_3 = (\mathrm{id}_{E_2 \otimes S_{2}^{}} \otimes \mathsf{SWAP}_{\overline{S_0^{\prime}}  R^{}})(\sigma_2).
\end{equation}
\item Take the local action of the intervention $\mathcal{V}^{\prime}$ with
\begin{equation}
\sigma_4= (\mathrm{id}_{E_2} \otimes \mathcal{V}^{\prime} \otimes \mathrm{id}_{\overline{S_0^{\prime}}})(\sigma_3).
\end{equation}
\item Finally, take the local action of the quantum channel $\mathcal{U}_4 \circ \mathcal{U}_3$ in order to have
\begin{equation}
\sigma_5=((\mathcal{U}_4 \circ \mathcal{U}_3) \otimes \mathrm{id}_{R^{\prime} \otimes \overline{S_0^{\prime}}})(\sigma_4).
\end{equation}
\end{enumerate}

Importantly, it is possible to compute the entropic terms in the right-hand-side (r.h.s.) of Eq.~(\ref{eq:thm_1}) from the $\sigma_5$ above.

\section{Alternative forms of the higher-order quantum Markov condition and the correspondent quantum information inequalities}\label{alternative_HOQMC}

Now, let us consider the possible non-trivial variations of the higher-order quantum Markov condition preserving its main property, that is, the maximum number of independent side-channels allowed in a multitime quantum process. Similarly to the presentation in Sec.~\ref{witness}, the focus here is on the case of three-slots quantum combs, since it is the simplest instance of the problem. 

\begin{figure*}[t] 
\centering   
\includegraphics[width=0.33\textwidth]{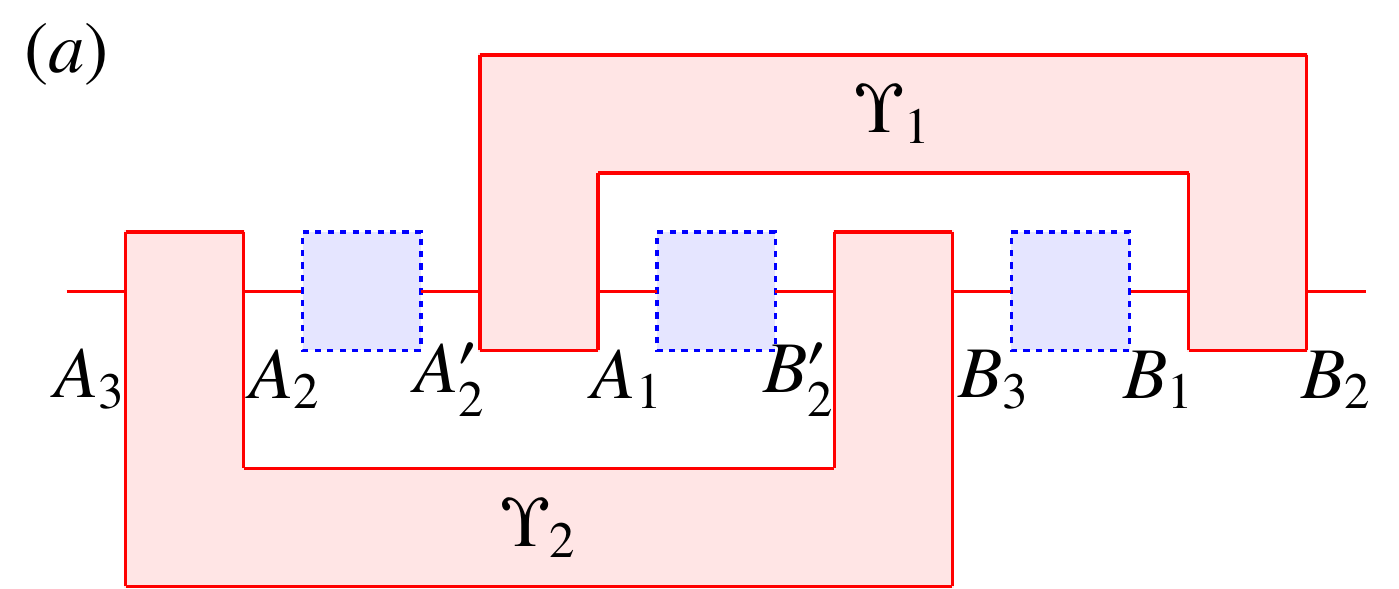}      
\includegraphics[width=0.33\textwidth]{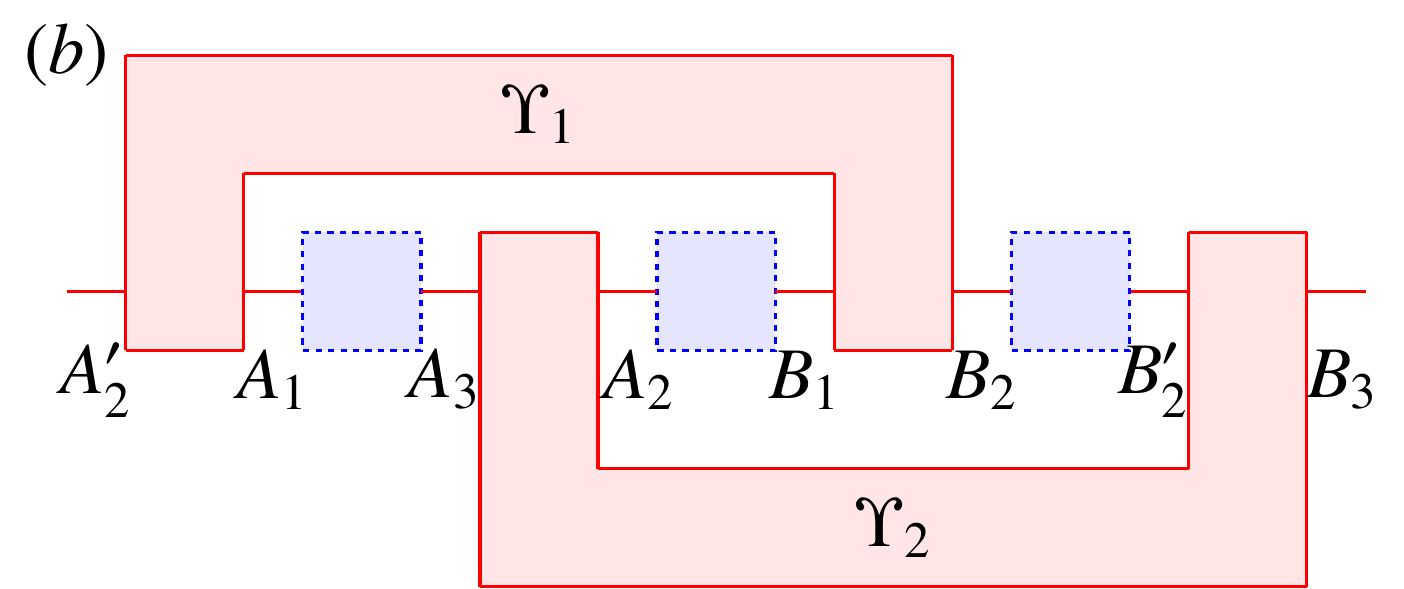} 
\includegraphics[width=0.33\textwidth]{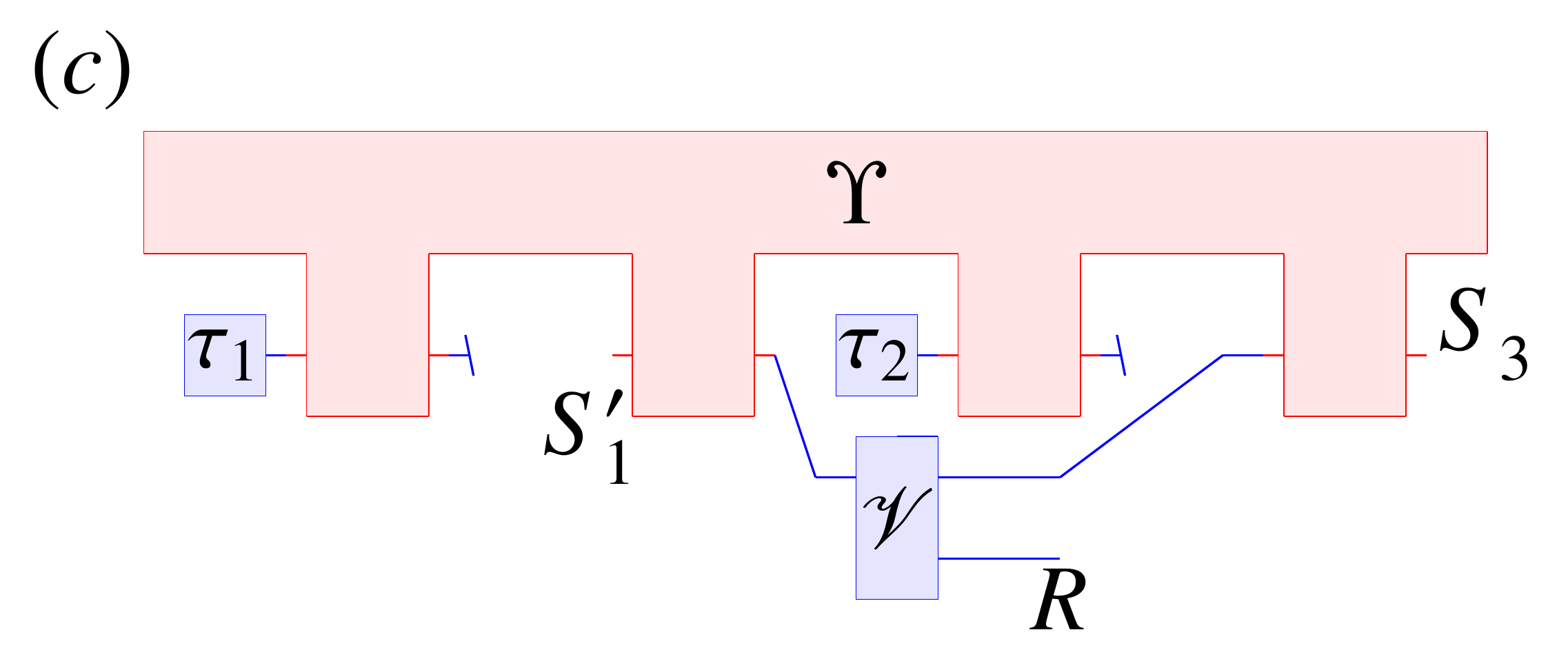}
\includegraphics[width=0.33\textwidth]{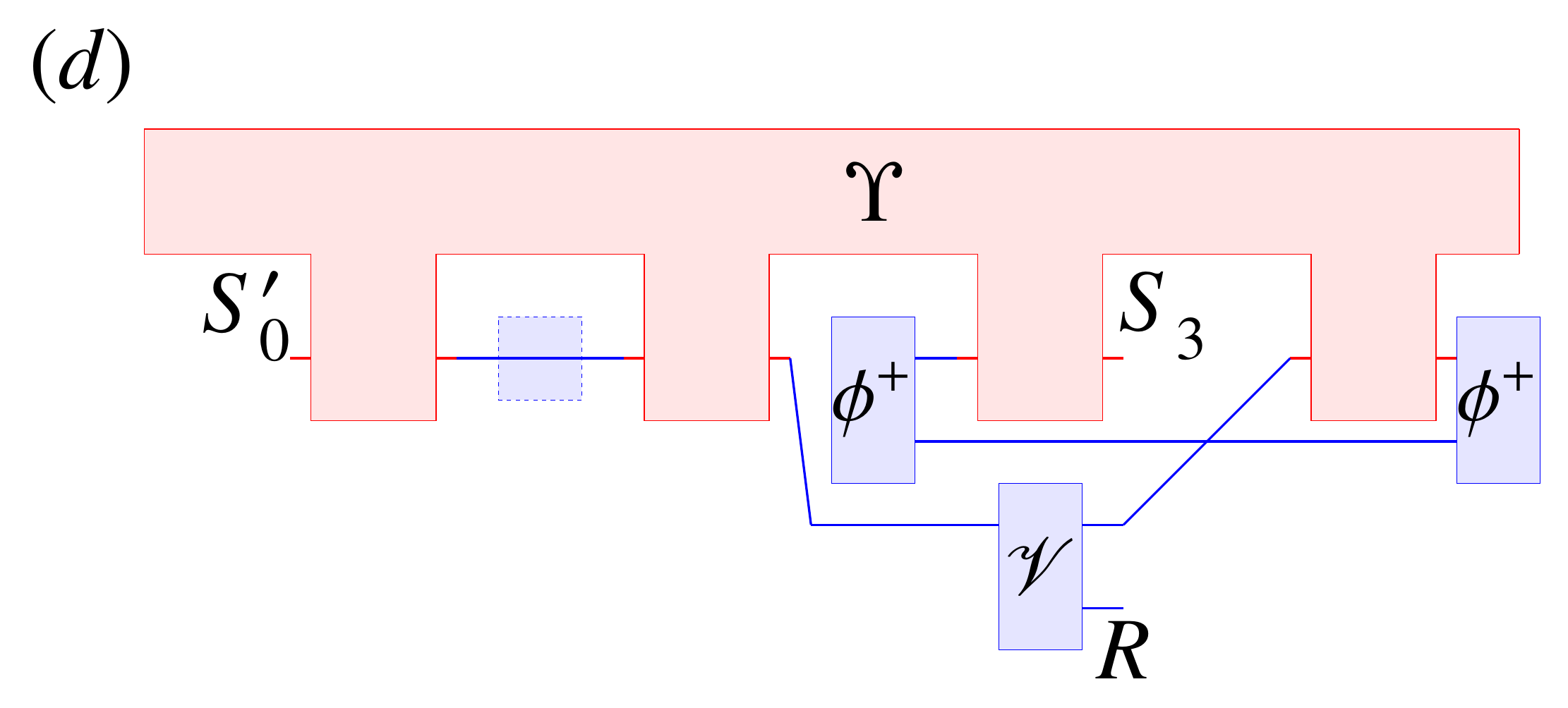}
\includegraphics[width=0.33\textwidth]{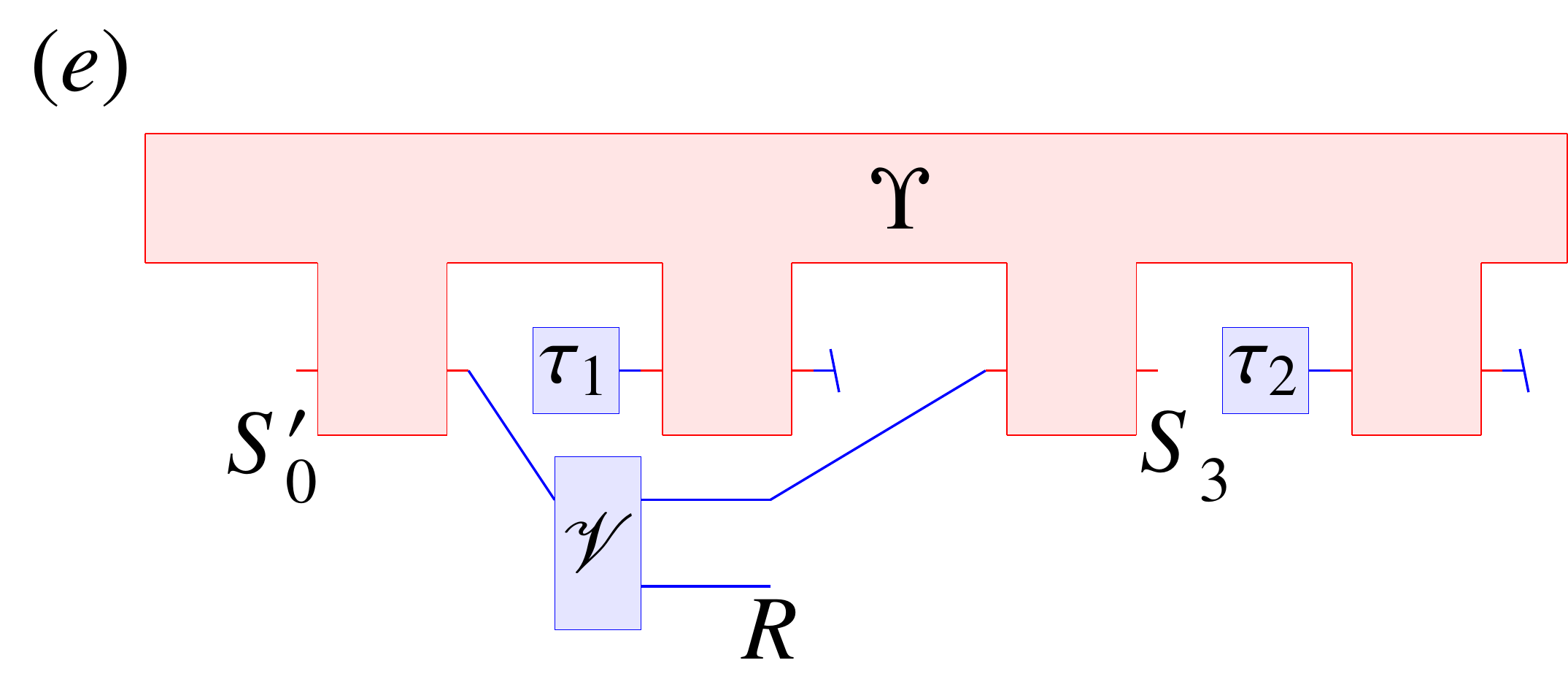}
\includegraphics[width=0.33\textwidth]{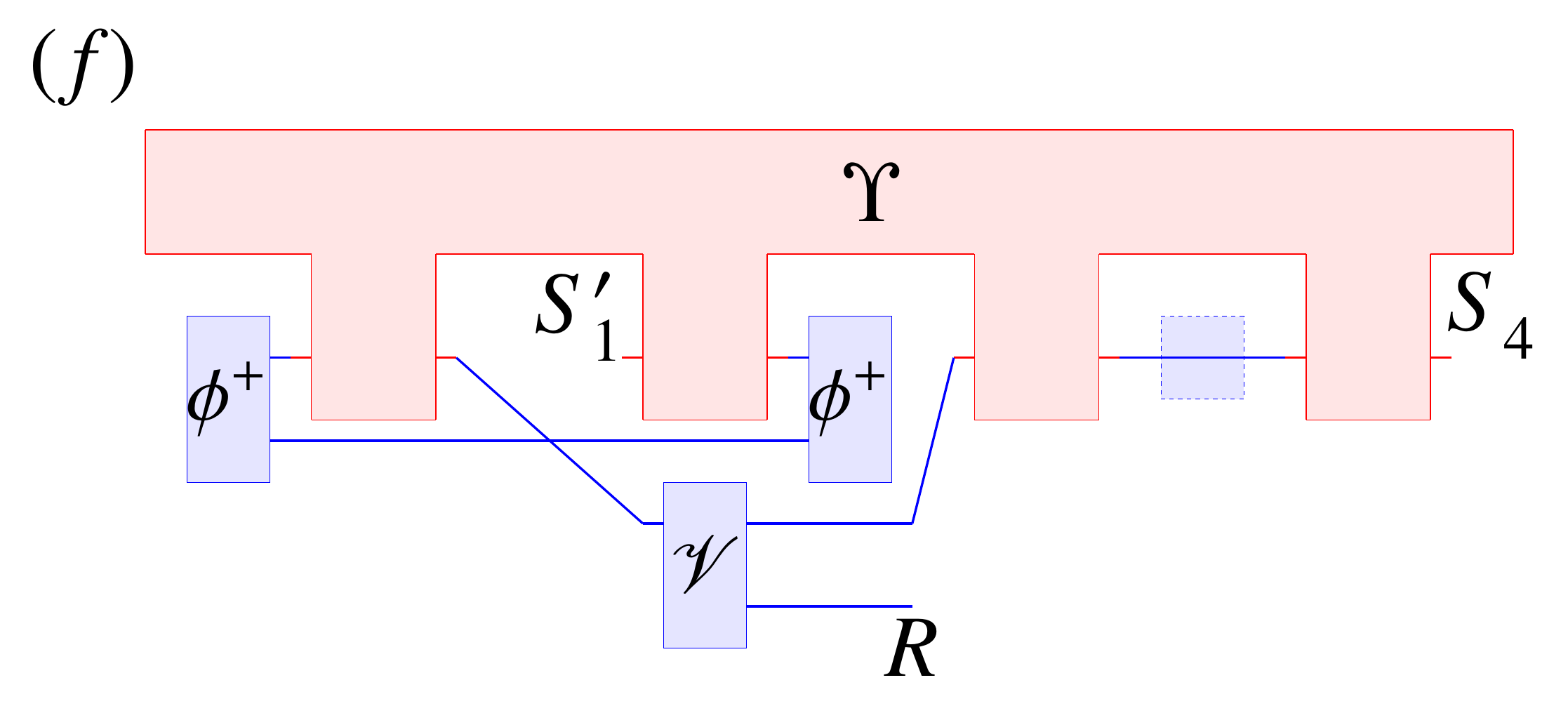}
\caption{Alternative form of the higher-order quantum Markov condition. Panels (a) and (b) show alternative forms of the standard model of quantum processes considered in this study. In turn, panels (c) and (d) provide diagrammatic representations of appropriate interventions for the validity of alternative information inequalities for the former, while panels (e) and (f) depict an intervention scheme for the later. In the pictures, the label of time instants follow the same convention as in Fig.\ref{fig:Sec_V_1}(a).}
    \label{fig:alternative_Markov}
\end{figure*}

In Figs.~\ref{fig:alternative_Markov}(a) and ~\ref{fig:alternative_Markov}(b), the two alternative versions of the model under consideration are represented in terms of diagrams.  In the former, the top-part of the quantum combs $\Upsilon_1$ and $\Upsilon_2$ are swapped, while in the later the bottom-part of the combs are swapped. Interestingly, it is possible to recover the desired outputs of the standard model from the intervention strategy depicted in Figs.~\ref{fig:alternative_Markov}(c) and~\ref{fig:alternative_Markov}(d), and Figs.~\ref{fig:alternative_Markov}(e) and~\ref{fig:alternative_Markov}(f), respectively. Hence, these intervention schemes clearly imply in alternative versions of the higher-order information inequality in Eq.~(\ref{eq:thm_1}), precisely stated as follows. 

\begin{lemma}[Reformulation of the results in Ref.~\cite{biamonte2017tensor}]\label{retrocausality_lemma}
Let $\mathcal{E}_{S_0\rightarrow S_1}^{(1)}$ and $\mathcal{E}_{S_1^{\prime}\rightarrow S_2}^{(2)}$ be quantum channels constituting pre- and post-processing operations of a  quantum Markov process $\Upsilon$, where the local quantum systems have the same dimension $d$. That is, for each input quantum channel $\mathcal{F}_{S_1 \rightarrow S_1^{\prime}}$ it holds that $\Upsilon(\mathcal{F})=\mathcal{E}^{(2)} \circ \mathcal{F} \circ \mathcal{E}^{(1)}$. Define the following intervention: $(i)$ Feed $S_0$ with part of a maximally entangled state $\phi_{S_0 \overline{S_0}}^{+}$; $(ii)$ Prepare the quantum system $S_{1}^{\prime}$ in some state $\rho$; $(iii)$ Measure the quantum system $S_2 \otimes \overline{S_0}$ in the state $\phi_{S_2 \overline{S_0}}^{+}$. Then, the result of such protocol is the preparation of the quantum state $(\mathcal{E}^{(1)}\circ \mathcal{E}^{(2)})(\rho)$ with probability $1/d^{2}$.
    
\end{lemma}
\begin{proof}
The steps $(i)$ and $(ii)$ of the intervention defined under the assumptions give the three-partite quantum state
\begin{equation}
    \sigma_{S_2\overline{S_0}S_1}=\mathcal{E}^{(2)}(\rho) \otimes  \left(\mathrm{id}_{\overline{S_0}} \otimes \mathcal{E}^{(1)}\right)(\phi^{+}) 
\end{equation}

Then let $\ket{i}$ denote a fixed orthonormal basis. The maximally entangled quantum state takes the form
\begin{equation}
\ket{\phi^{+}}_{\overline{S_0}S_0}=\frac{1}{\sqrt{d}}\sum_{i=1}^{d} \ket{i}_{\overline{S_0}} \otimes \ket{i}_{S_0}.
\end{equation}

In turn, the quantum state preparation in item $(i)$ for $S_1^{\prime}$ can be expressed as
\begin{equation}
    \rho_{S_1^{\prime}}=\sum_{k,l=1}^{d} \rho_{k,l} \ket{k} \bra{l}_{S_1^{\prime}}
\end{equation}

Hence, we must have
\begin{equation}
    \sigma=\frac{1}{d} \sum_{\substack{i,j\\ k,l}} \rho_{k,l} \mathcal{E}^{(2)}(\ket{k}\bra{l}) \otimes \left[ \ket{i}\bra{j}  \otimes \mathcal{E}^{(1)}(\ket{i}\bra{j}) \right]
\end{equation}

The post-measurement quantum state $\sigma^{\prime}$ defined under the intervention item $(iii)$ with probability of occurrence denoted as $p$, for the measurement operator $M_{S_2 \overline{S_0} S_1}=\phi_{S_2 \overline{S_0}}^{+}\otimes \mathds{1}_{S_1}$, is given by
\begin{widetext}
\begin{eqnarray}
    \sigma^{\prime}&=&\frac{M \rho M^{\dagger}}{p} \\
    &=&\frac{1}{p}\left( \frac{1}{d} \sum_{m,n} \ket{m,m}\bra{n,n} \otimes \mathds{1} \right) 
 \cdot \left( \frac{1}{d} \sum_{\substack{i,j\\ k,l}} \rho_{k,l} \mathcal{E}^{(2)}(\ket{k}\bra{l}) \otimes \left[ \ket{i}\bra{j}  \otimes \mathcal{E}^{(1)}(\ket{i}\bra{j}) \right] \right) \cdot
 \left( \frac{1}{d} \sum_{r,s} \ket{r,r}\bra{s,s} \otimes \mathds{1} \right)
 \\
 &=&  \frac{1}{p}\frac{1}{d^{3}}
 \sum_{\substack{i,j\\ k,l \\ m \\s}} \left[ \rho_{k,l}  \bra{n}  \mathcal{E}^{(2)}(\ket{k}\bra{l}) \ket{r} \ket{m} \bra{s} 
 \right] \otimes \left[  \bra{n}\ket{i} \bra{j}\ket{r} \ket{m} \bra{s} \right] \otimes  \mathcal{E}^{(1)}(\ket{i}\bra{j}).
\end{eqnarray}
where it has been used the notation $\ket{i}\otimes \ket{j}=\ket{i,j}$.

Thus, we have the reduced state preparation for $S_1$ given by
\begin{eqnarray}
\sigma_{S_0}^{\prime}&=&\Tr_{S_2 \overline{S_0}} \left[\sigma_{S_2 \overline{S_0} S_1}^{\prime}\right] \\
&=& \frac{1}{p}\frac{1}{d^{3}}
 \sum_{\substack{i,j\\ k,l \\ m}} \rho_{k,l}  \bra{i}  \mathcal{E}^{(2)}(\ket{k}\bra{l}) \ket{j} \mathcal{E}^{(1)}(\ket{i}\bra{j}) \\
&=&  \frac{1}{p}\frac{1}{d^{3}}
 \sum_{\substack{i,j \\ m}}   \bra{i}  \mathcal{E}^{(2)}\left( \sum_{k,l}\rho_{k,l}\ket{k}\bra{l}\right) \ket{j} \mathcal{E}^{(1)}(\ket{i}\bra{j}) \\
 &=&  \frac{1}{p}\frac{1}{d^{3}}
 \sum_{\substack{i,j \\ m}}   \bra{i}  \mathcal{E}^{(2)}\left( \rho \right) \ket{j} \mathcal{E}^{(1)}(\ket{i}\bra{j}) \\
 &=&  \frac{1}{p}\frac{1}{d^{3}}
 \sum_{m} \mathcal{E}^{(1)}\left(\sum_{\substack{i,j }}   \bra{i}  \mathcal{E}^{(2)}\left( \rho \right) \ket{j} \ket{i}\bra{j}\right) \\
 &=& \frac{1}{p}\frac{1}{d^{2}}
 \mathcal{E}^{(1)} \left( \mathcal{E}^{(2)} ( \rho) \right), 
\end{eqnarray}
and furthermore, we must also have $p=1/d^2$. 
\end{widetext}
\end{proof}

The quantum intervention described above is sometimes called retrocausal communication \cite{ji2026retrocausal}, as the result of such intervention is to make the output of $\mathcal{E}^{(2)}$ under the quantum state preparation $\rho$ to be the input of $\mathcal{E}^{(1)}$. That is, the quantum communication goes backwards in time.
Unfortunately, the post-selection stage is unlikely to take place in very high-dimensional quantum systems as $p \rightarrow 0$ as $d \rightarrow \infty$. Nevertheless, it is a real possibility for the qubit systems considered in our examples, with $p=1/4$.

Importantly, as a consequence of Lemma~\ref{retrocausality_lemma} along with the standard formulation of the higher-order quantum information inequality in Corollary~\ref{thm_1}, it follows precise information-theoretic bounds on the alternative versions of higher-order Markovianity.

\begin{corollary}[Alternative versions of the higher-order quantum information inequality]\label{alternative_HOQII}Let $\Upsilon$ be a five-steps quantum process for which holds either the condition in Fig.~\ref{fig:alternative_Markov}(a) or Fig.~\ref{fig:alternative_Markov}(b).  Then, it holds the information inequalities
\begin{multline}\label{eq:alternative_HOQII_1}
H(S_1^{\prime} \rightarrow S_4^{})_{\Tr_{R}\circ \Omega_1^{\prime}}-H(S_1^{\prime} \rightarrow S_4^{} \otimes R)_{\Omega_1^{\prime}} \\
\geq H(S_0^{\prime} \rightarrow S_3^{})_{\Tr_R \circ \Omega_2^{\prime}}-H(S_0^{\prime} \rightarrow S_3^{} \otimes R)_{\Omega_2^{\prime}},
\end{multline} 
and 
\begin{multline}\label{eq:alternative_HOQII_2}
H(S_0^{\prime} \rightarrow S_3^{})_{\Tr_{R}\circ \Omega_1^{\prime\prime}}-H(S_0^{\prime} \rightarrow S_3^{} \otimes R)_{\Omega_1^{\prime\prime}} \\
\geq H(S_1^{\prime} \rightarrow S_4^{})_{\Tr_R \circ \Omega_2^{\prime\prime}}-H(S_1^{\prime} \rightarrow S_4^{} \otimes R)_{\Omega_2^{\prime\prime}},
\end{multline}
respectively. Here, $\Omega_1^{\prime}$ and $\Omega_2^{\prime}$ are the quantum channels defined with respect to the diagrams in Figs.~\ref{fig:alternative_Markov}(c) and~\ref{fig:alternative_Markov}(d), respectively. On the other hand, the quantum channels $\Omega_1^{\prime\prime}$ and $\Omega_2^{\prime\prime}$ are defined with respect to the diagrams in Figs.~\ref{fig:alternative_Markov}(e) and~\ref{fig:alternative_Markov}(f), respectively.
\end{corollary}

As mentioned previously, the intervention strategy adopted in the result above is done in order to reproduce the same outputs as expected by the standard form of higher-order Markovianity. It is worth to mention that the main ingredient here is the preparation of the quantum system $S_{t}^{\prime}$ and some auxiliary system $\overline{S_{t}^{\prime}}$ in a maximally entangled quantum state $\phi^{+}$, and take a post-selection of the system $S_{t+2}^{out}$ and the auxiliary system in the same maximally entangled state, with $t=0,2$. This procedure forces the alternative versions of the model to be found in its standard form \cite{biamonte2017tensor}. Furthermore, the proof of Corolary~\ref{alternative_HOQII} follows directly from Corolary~\ref{thm_1}: The later is a simple reformulation of the former. On another side, such post-selection might not be easily found---with non-vanishing probability---in an unknown quantum process one would be interested to provide a characterization of its non-Markovian effects. 

In order to assess how the alternative versions of the higher-order quantum information inequality work in practice, we study how the standard higher-order Markov model might violate the information inequalities emerging from its alternative forms. Note that for this case the post-selection procedure can be done with non-vanishing probability of success. Now, let us consider the violation of the alternative quantum information inequalities for a standard model with unitary blocks consisting of the unitary quantum operation defined in Eq.~(\ref{unitary_matrix}), thus defining a range of parameters $0 \leq  \eta \leq 1$.

\begin{figure}[h] 
\centering   
\includegraphics[width=0.23\textwidth]{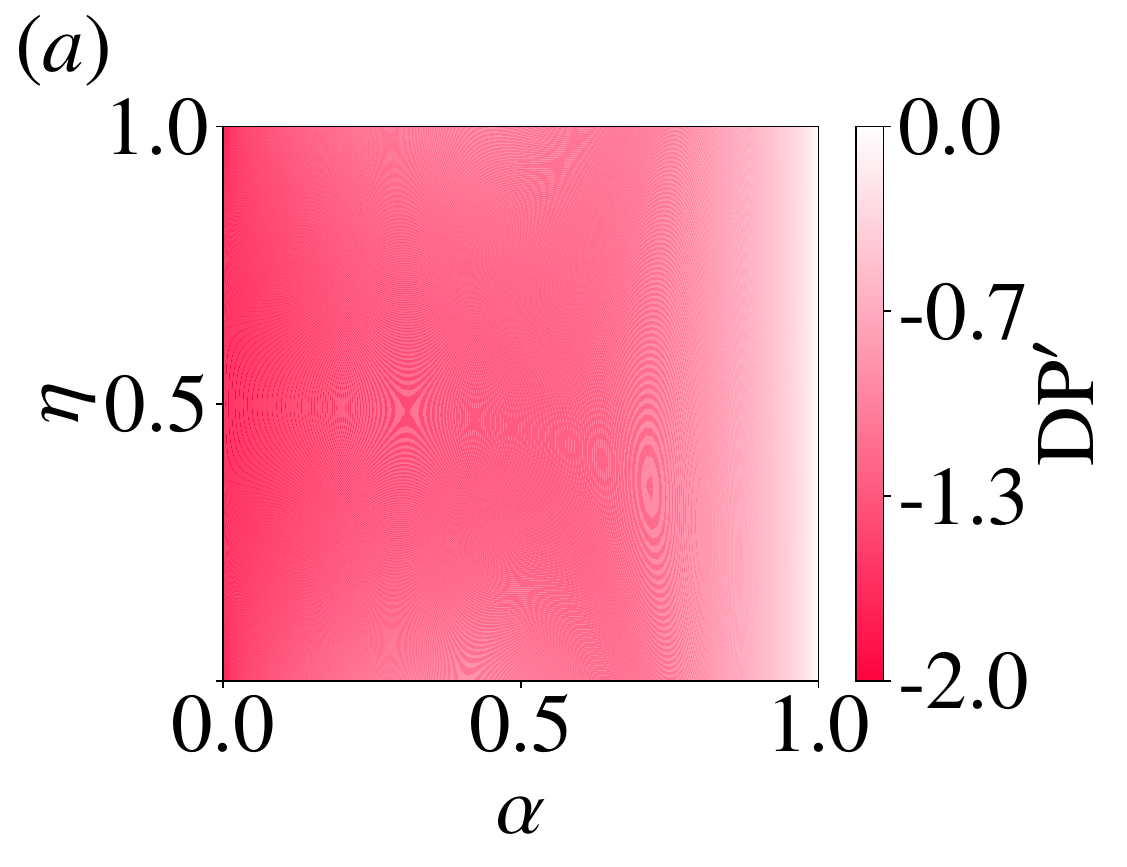} 
\includegraphics[width=0.23\textwidth]{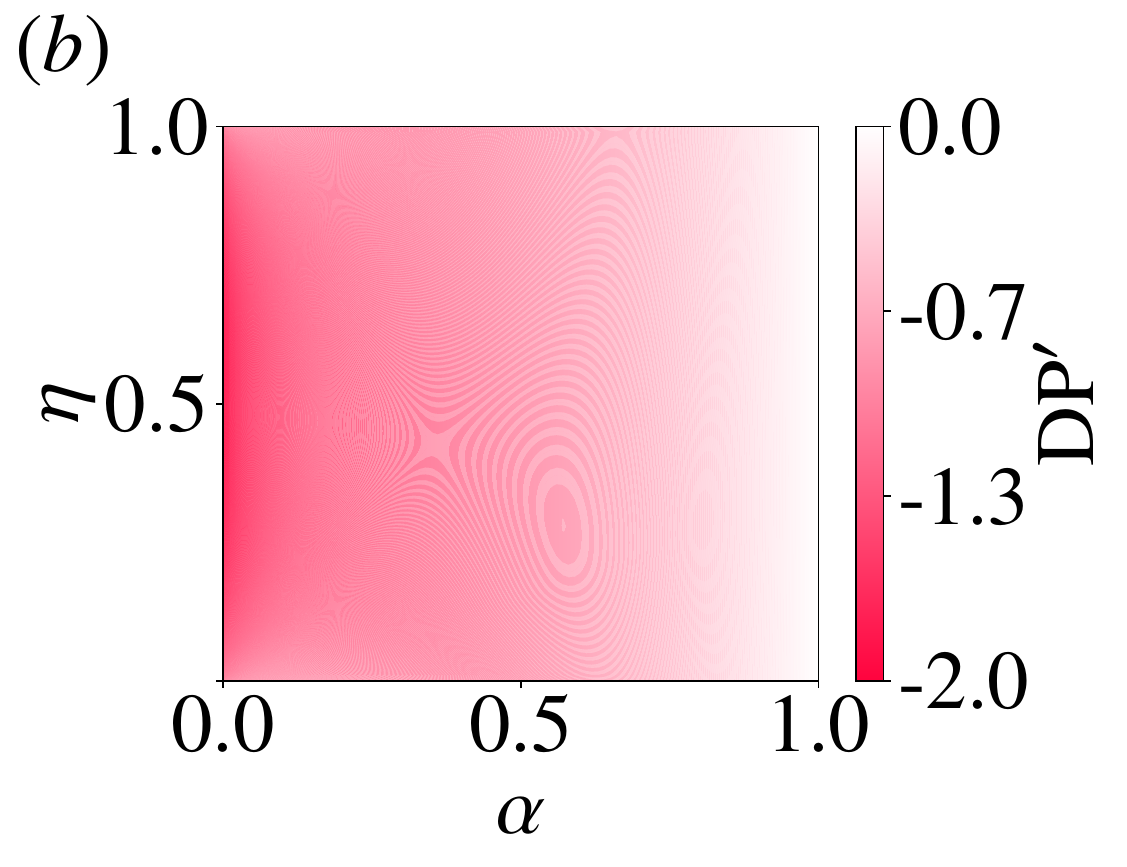} 
\includegraphics[width=0.23\textwidth]{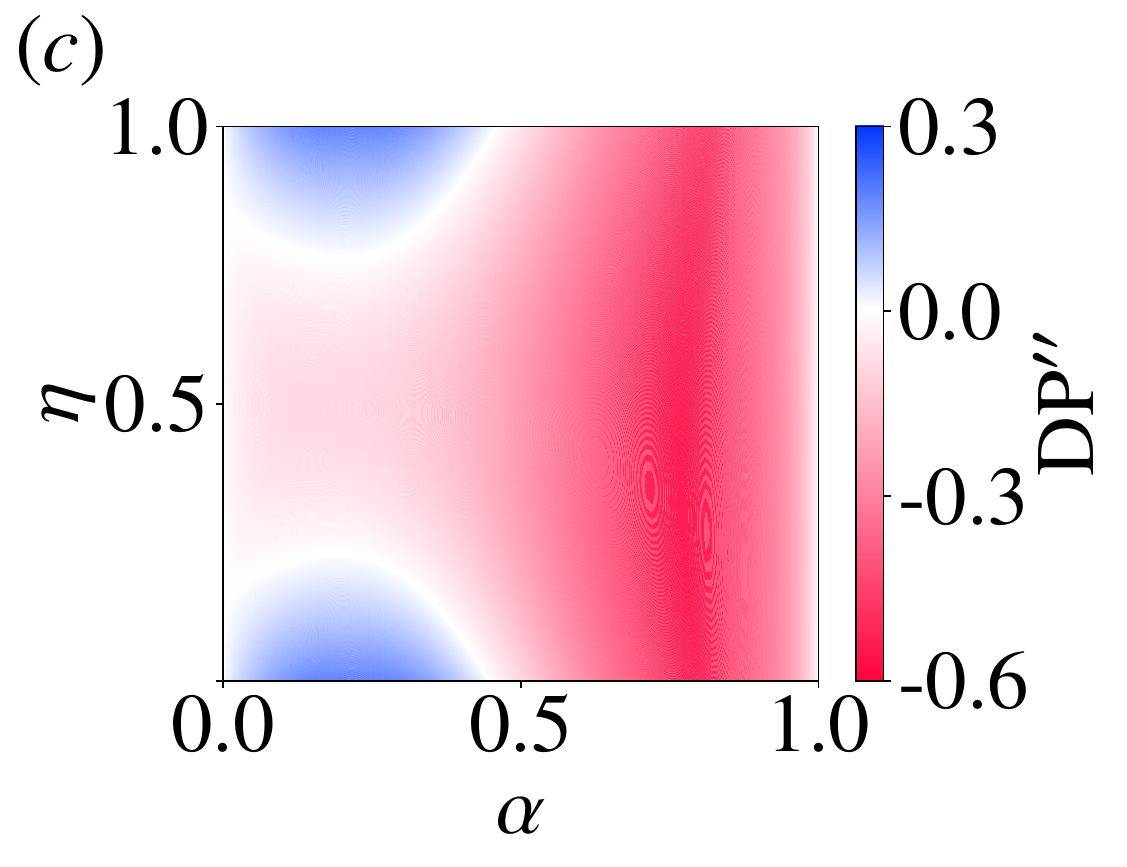} 
\includegraphics[width=0.23\textwidth]{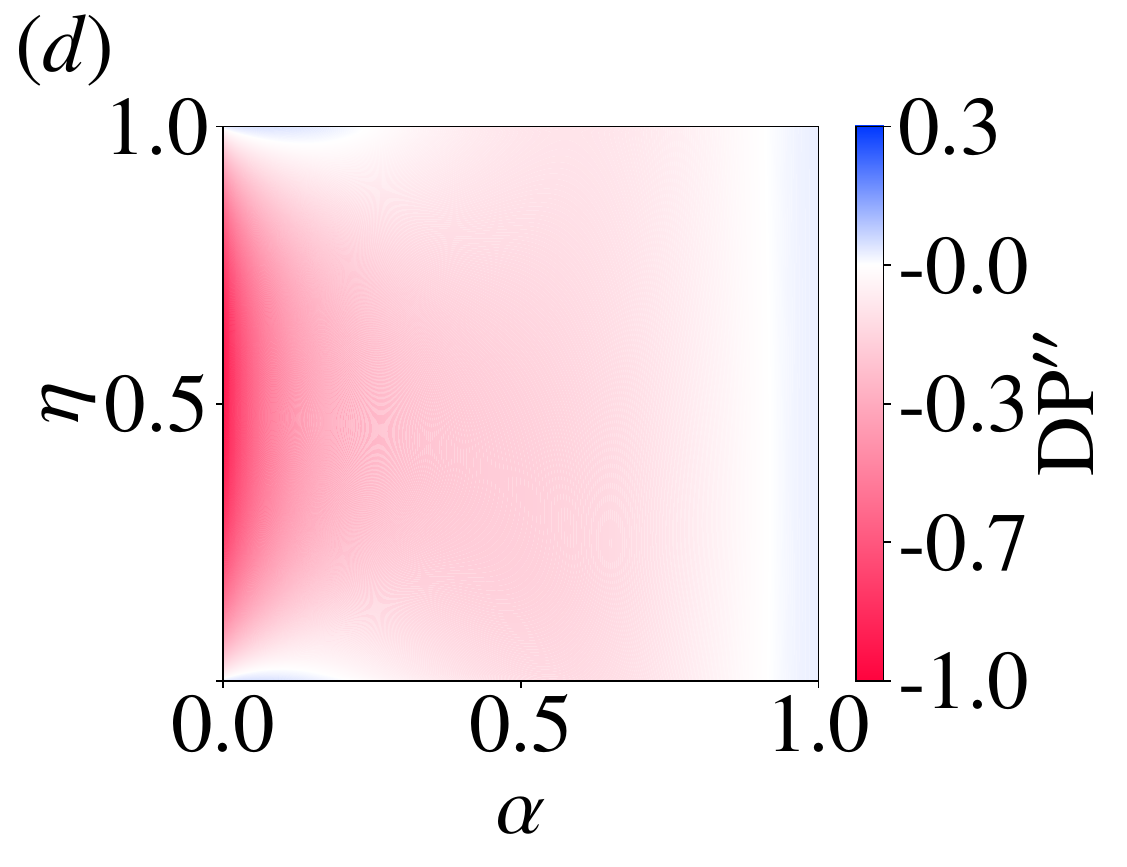} 

\caption{Violation of the alternative higher-order quantum information inequalities. The panels (a) and (b) show the gap in the violation of Eq.~(\ref{eq:alternative_HOQII_1}) for the initial environment state preparations $\ket{0}$ and $\ket{+}$ denoted with $\mathrm{DP}^{\prime}$, respectively, while panels (c) and (d) present the gap in the violation of Eq.~(\ref{eq:alternative_HOQII_2}) denoted with $\mathrm{DP}^{\prime\prime}$ for the same respective cases.}
    \label{fig:alternative_HOQII}
\end{figure}

Figure~\ref{fig:alternative_HOQII} consists of the study of the violation of Eqs.~(\ref{eq:alternative_HOQII_1}) and (\ref{eq:alternative_HOQII_2}) by a standard higher-order Markov model defined with unitary quantum channels as in Eq.~(\ref{unitary_matrix}), where each separate quantum comb, $\Upsilon_1$ and $\Upsilon_2$ in Fig.~\ref{figure1}(a), have the environment system initially prepared on the same quantum state $\varphi$. The intervention isometry $\mathcal{V}$ is set as the same keep-and-prepare quantum channel as before, thus defining a range of parameters $0 \leq  \alpha \leq 1$. Figs.~\ref{fig:alternative_HOQII}(a) and ~\ref{fig:alternative_HOQII}(b) show the violation of Eq.~(\ref{eq:alternative_HOQII_1}) for initial environment state preparations $\ket{0}$ and
$\ket{+}$, respectively. In turn, Figs.~\ref{fig:alternative_HOQII}(c) and ~\ref{fig:alternative_HOQII}(d) respectively presents the violation of Eq.~(\ref{eq:alternative_HOQII_2}) for the same initial environment state preparations as in the previous case. Importantly, the translation of the process diagrams in Fig.~\ref{fig:alternative_Markov} into simple quantum circuit computations is analogous to the presentation in Appendix~\ref{appendix_intervention}.

For the particular example addressed here, the quantum information inequality emerging from the alternative higher-order Markov condition in Fig.~\ref{fig:alternative_Markov}(a) is violated for almost every value of the parameters $\eta,\alpha$. See top panels (a) and (b) in Fig.~\ref{fig:alternative_HOQII}. On another side, the complementary quantum information inequality, emerging from processes of the form presented in Fig.~\ref{fig:alternative_Markov}(b), highly depends on the values of $\eta,\alpha$. See bottom panels (c) and (d) in Fig.~\ref{fig:alternative_HOQII}. It is worth to note that there is no straightforward reason for which the condition in Eq.~(\ref{eq:alternative_HOQII_1}) is more efficient in the detection of non-Markovian effect than its complementary condition in Eq.~(\ref{eq:alternative_HOQII_2}). Nevertheless, a more complete investigation of this phenomena is left to further work.

\section{Higher-order quantum Markov monogamy inequalities}\label{appendix_Markov_monogamy}

\begin{figure*}[t] 
\centering   
\includegraphics[width=0.7\textwidth]{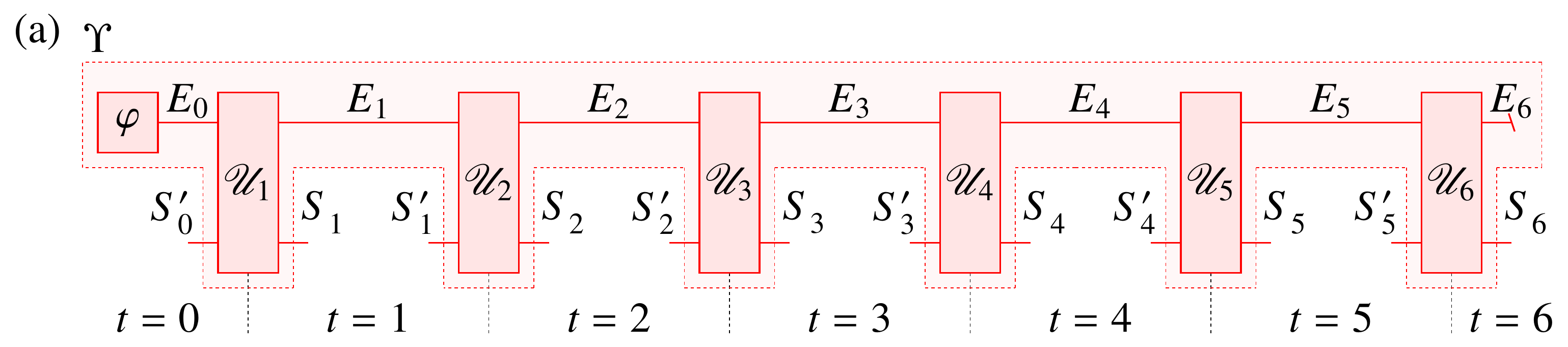}      
\includegraphics[width=0.7\textwidth]{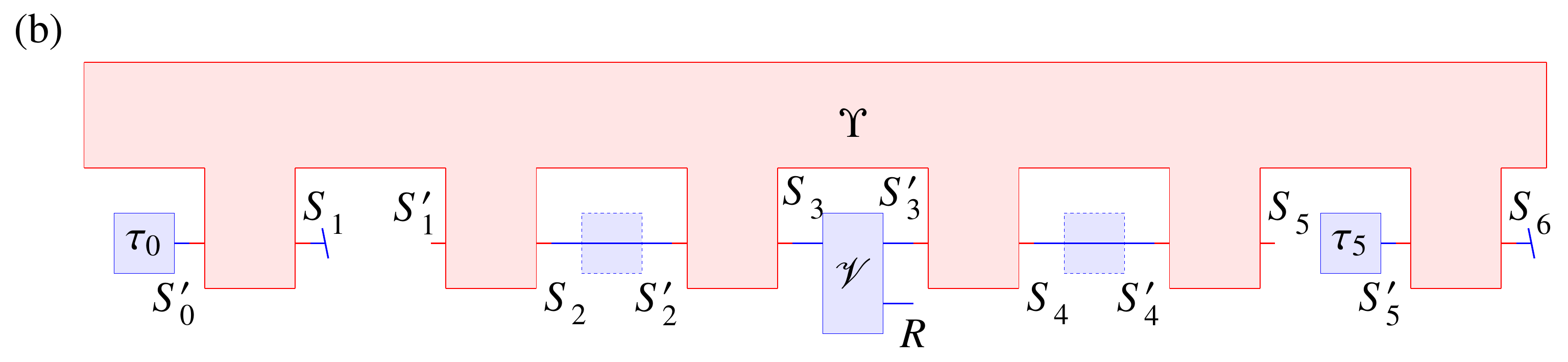} 
\includegraphics[width=0.7\textwidth]{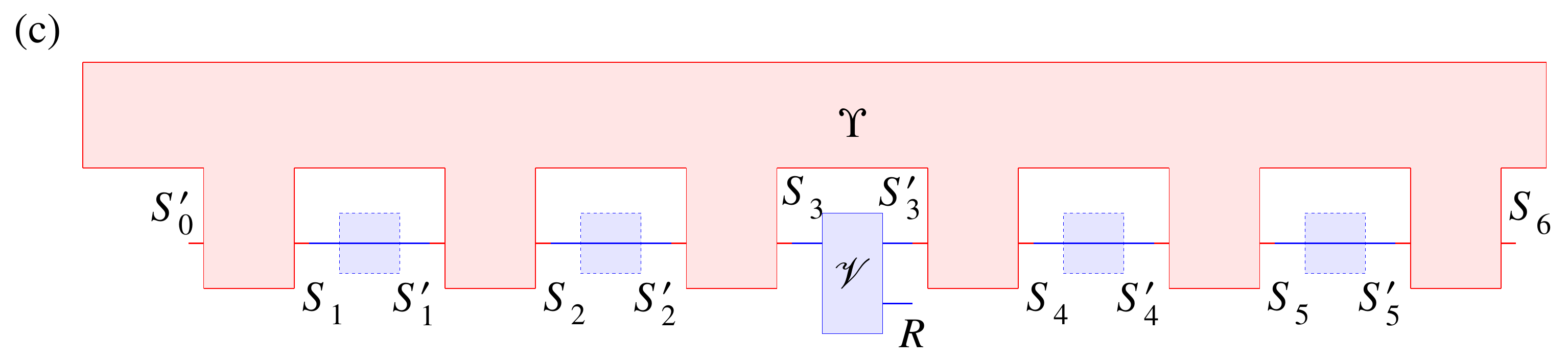}
\includegraphics[width=0.7\textwidth]{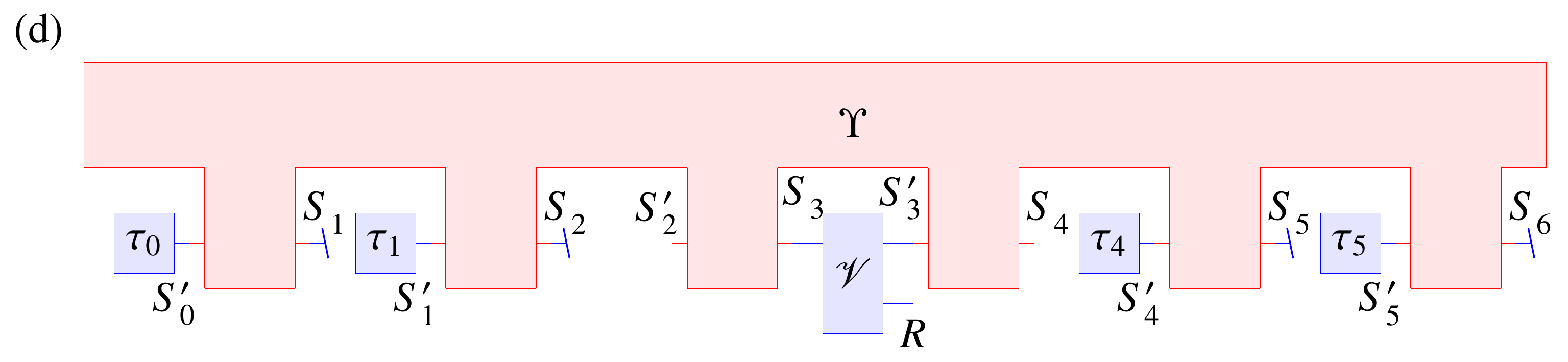}
\includegraphics[width=0.7\textwidth]{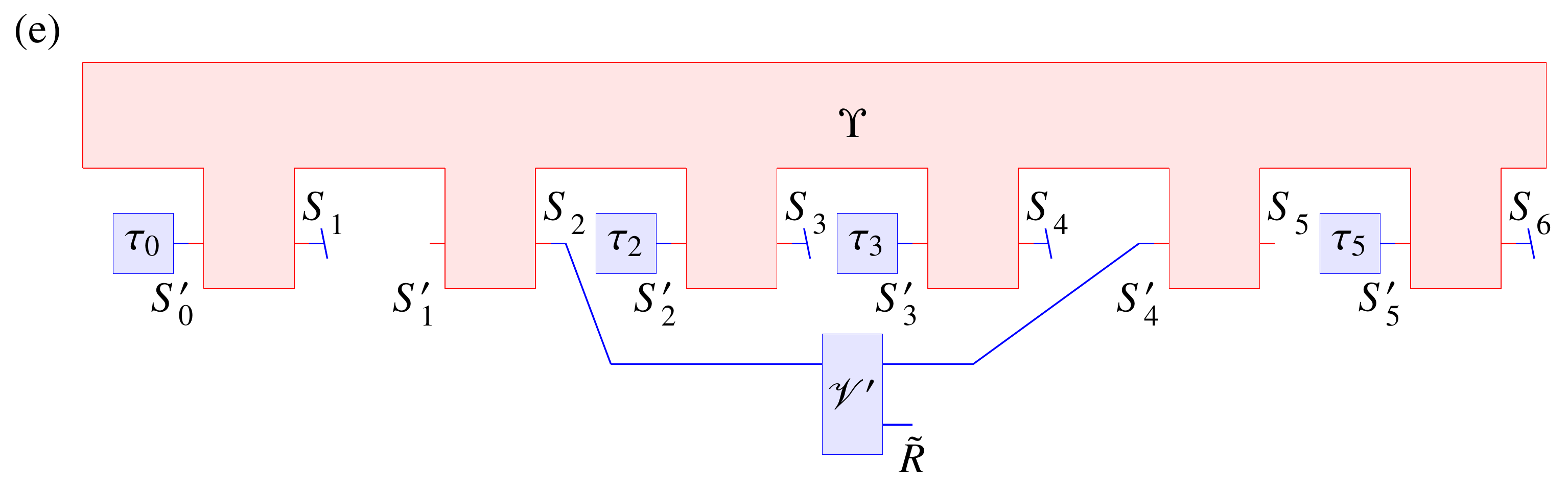}
\includegraphics[width=0.7\textwidth]{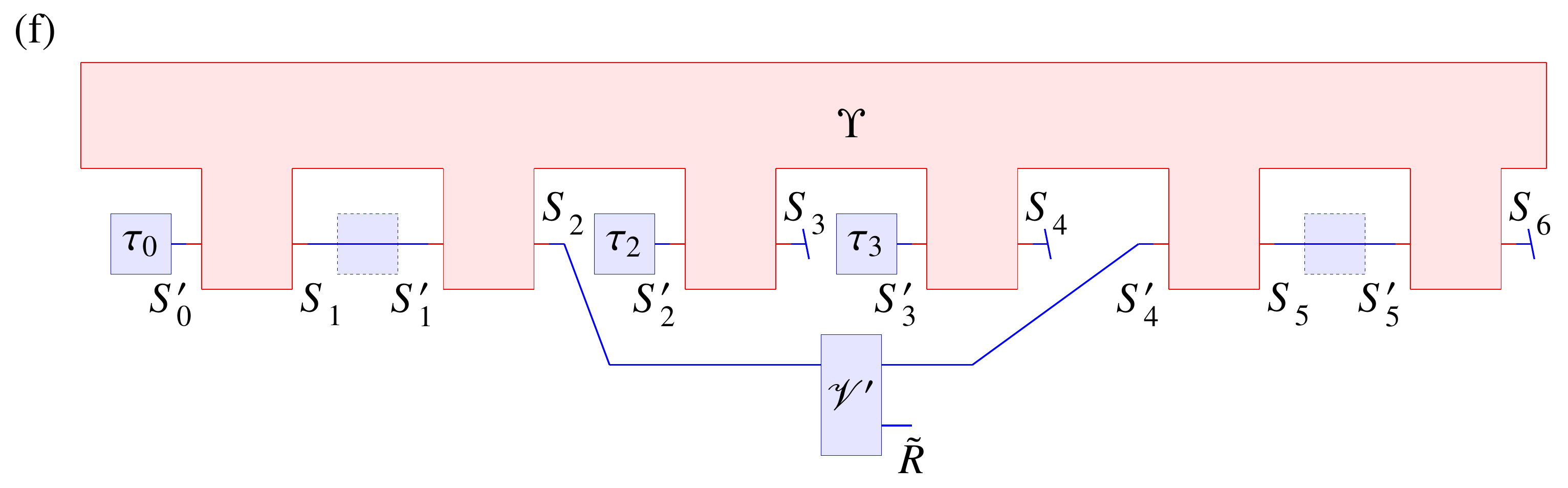}
\includegraphics[width=0.7\textwidth]{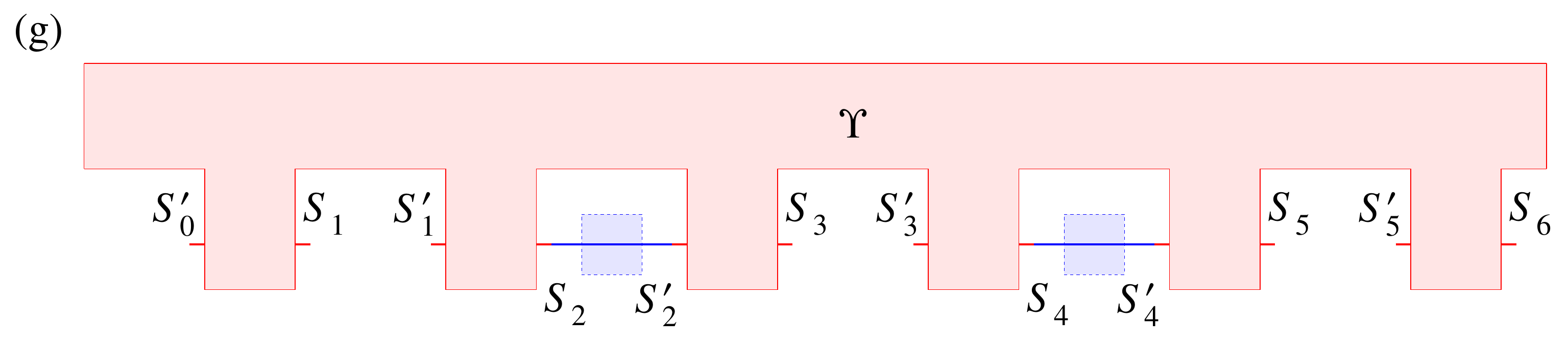}
\includegraphics[width=0.7\textwidth]{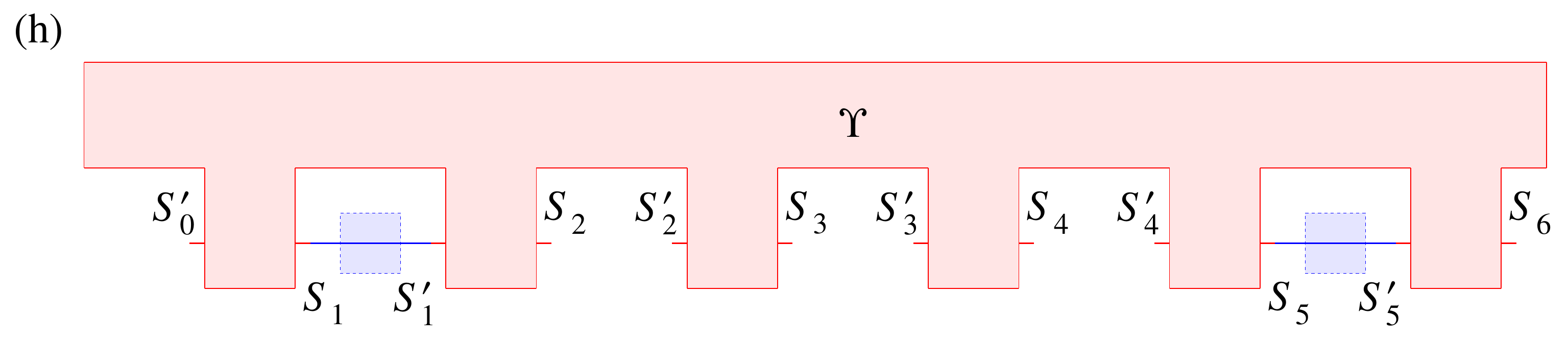}
\caption{Seven-steps quantum processes and the intervention strategy for the higher-order quantum Markov monogamy inequality. Panel (a) and (b) depict the diagram of an arbitrary seven-steps quantum process. Panels (b) and (c) show the interventions $\mathcal{I}_1$ and $\mathcal{I}_1^{\prime}$, respectively. Panels (d) and (e) show the interventions $\mathcal{I}_2$ and $\mathcal{I}_2^{\prime}$, respectively. Finally, panels (g) and (h) provide the diagrammatic representations for reducing seven-steps quantum processes in five-steps ones preserving the higher-order Markov property.}
    \label{fig:HOQMMI}
\end{figure*}

As discussed in Section~\ref{inf_ineqs_section}, classical information inequalities provide an appropriate pattern for the establishment of quantum information inequalities. Nevertheless, in general, not every limitation of the former kind implies in a valid bound of the later. Indeed, that is the case for the condition in Eq.~(\ref{dpi_2}), which holds in quantum information-processing only for isometric quantum channels in general. 

Interestingly, when classical processes with more than three-steps are under consideration, a distinct form of classical information inequalities emerges. For instance, it holds the classical Markov monogamy inequality for each four-step classical Markov process \cite{capela2020monogamy,yeung2008information}, 
\begin{equation}
 I(X_1:X_4)+I(X_2:X_3) \geq I(X_1:X_3)+I(X_2:X_4).   
\end{equation}

Note that the information inequality above is time-symmetric, and has a valid quantum version \cite{capela2022quantum}, 
\begin{multline}
    I_c(\rho ; \Lambda_3 \circ \Lambda_2 \circ \Lambda_1) + I_c(\Lambda_1(\rho) ; \Lambda_2) \\ \geq I_c(\Lambda_1(\rho) ; \Lambda_3 \circ \Lambda_2) +  I_c(\rho ; \Lambda_2 \circ \Lambda_1),
\end{multline}
where $\rho$ is an arbitrary initial quantum state, and $\Lambda_1,\Lambda_2,\Lambda_3$ are arbitrary quantum channels (up to appropriate input and output dimensions). 

The quantum Markov monogamy inequality above was conjectured to hold for an arbitrary number of quantum channels (that is, for quantum Markov processes with any number of steps), and proven to be valid for a number of steps up to eight. A proof for quantum Markov processes with arbitrary number of steps still remains as an open problem.

Following the reasoning in Appendix~\ref{appendix_proof}, one is able to prove that each quantum information inequality imply a correspondent higher-order quantum information inequality. Thus, it holds the higher-order quantum Markov monogamy inequality as stated in the following.
\begin{theorem}[Higher-order quantum Markov monogamy inequality]\label{thm:HOQMMI} Let $\Lambda$ be a quantum channel and $\Upsilon_1,\Upsilon_2,\Upsilon_3$ be single-slot quantum combs (with dimensions for the local systems such that each term of the inequality below are well-defined). Then, it holds that
\begin{multline}\label{HOQMMI}
    \widetilde{I_c}(\Lambda ; \Upsilon_3 \circ \Upsilon_2 \circ \Upsilon_1) + \widetilde{I_c}(\Upsilon_1(\Lambda) ; \Upsilon_2) \geq \\
     \widetilde{I_c}(\Upsilon_1(\Lambda) ; \Upsilon_3 \circ \Upsilon_2) +  \widetilde{I_c}(\Lambda ; \Upsilon_2 \circ \Upsilon_1).
\end{multline} 
\end{theorem}

The proof for the result above is similar to the presented in Appendix~\ref{appendix_proof}, and thus, it is omitted here. The interested reader is also urged to follow Ref.~\cite{capela2022quantum} for a detailed discussion on the original information inequality.

Importantly, the condition in Eq.~(\ref{HOQMMI}) can be stated following an interventional perspective as well. For that sake, let $\Upsilon$ be a five-slot quantum comb, that is, a seven-steps quantum process (see Fig.~\ref{fig:HOQMMI}(a) for a diagrammatic representation). Here, our ultimate goal would be to find a condition corresponding to Theorem~\ref{thm:HOQMMI} in such a way that from its violation one makes sure a given process is not compatible with higher-order Markovianity from local interventions only. That is, to be able to prove the process as a non-sequential action of single-slot quantum combs $\Upsilon \neq \bigcirc_{k=1}^{3} \Upsilon_k$ (similar note as in \cite{Note1} in regard to the product form of the process matrix representation applies to this case as well). 

Equation~(\ref{HOQMMI}) consists of an inequality with respect to four different (higher-order) coherent information terms. In the interventional approach, we are given with an unknown seven-steps quantum process $\Upsilon$ for which we aim to decide on its non-compatibility with the higher-order Markov condition, and each term can be computed from a particular recipe of quantum operations acting on the process. 

Similarly as done in Appendix~\ref{appendix_intervention}, let us now define each intervention for the computation of the desired information quantities. The reader is urged to follow the precise definition that follows in the main text by consulting the diagrams in Fig.~\ref{fig:HOQMMI}. It is worth to note that this particular choice of intervention suffices for witnessing the incompatibility with higher-order quantum Markovianity, although it is not the unique option. That is, given an unknown quantum processes, acting on it with local quantum operations do not increase time-correlations, and thus, local channels can be fixed on the interventions in such a way that it would be absorbed by the process as a resulting overall effect.

The intervention $\mathcal{I}_1$ (see Fig.~\ref{fig:HOQMMI}(b)) is implemented in order to make possible the computation of the term $\widetilde{I_c}(\Lambda ; \Upsilon_2 \circ \Upsilon_1)$, and it is defined with the following collection of quantum operations:
\begin{enumerate}[$(i)$]
    \item First, make fixed any (isometric) quantum channel $\mathcal{V}$ from $S_{3}^{}$ to $S_{3}^{\prime} \otimes R$, so that it would have the same role of the isometric dilation of the input quantum channel $\Lambda$.
    \item Feed the slots at $t=2$ and $t=4$ with identity quantum channels. To be precise, there is some freedom on this part of the procedure: It would be also valid to make fixed any local quantum channel on each slot, and thus, to be considered as absorbed by the unknown process. That would be an interesting approach for when the input and output local systems have different dimensions. Here, the identity channels are adopted, since it is the simplest quantum channel to implement locally whenever possible. That is, without lost of generality, we consider the local systems to have the same dimension. 
    \item Feed the local output quantum system at steps $t=0$ and $t=5$ with arbitrarily fixed quantum states $\tau_1$ and $\tau_2$, respectively. 
    \item Finally, discard the local input quantum systems at steps $t=1$ and $t=6$. The remaining quantum operation is a quantum channel from which it is possible to compute the desired quantum information quantity.
\end{enumerate}

Somewhat similar is the following definition of the intervention $\mathcal{I}_1^{\prime}$ (see Fig.~\ref{fig:HOQMMI}(c)), from which it is possible to compute the term $\widetilde{I_c}(\Lambda ; \Upsilon_3 \circ \Upsilon_2 \circ \Upsilon_1)$:
\begin{enumerate}[(i)]
    \item Repeat the items (i) and (ii) from the definition of the intervention $\mathcal{I}_1$.
    \item Finally, feed the steps $t=1$ and $t=5$ with identity quantum channels. The output quantum operation is a quantum channel from which it is possible to compute the desired quantum information measure.
\end{enumerate}

Now, we can set the intervention $\mathcal{I}_2$ for the computation of the term $\widetilde{I_c}(\Upsilon_1(\Lambda) ; \Upsilon_2)$. In this stage, it is relevant to make fixed any isometric dilation $\mathcal{V}^{\prime}$ of the quantum channel from $S_{2}^{\prime}$ to $S_{4}^{}$ defined in the diagram of Fig.~\ref{fig:HOQMMI}(d). Then, the intervention is defined with the following protocol (see Fig.~\ref{fig:HOQMMI}(e)).
\begin{enumerate}[$(i)$]
    \item Take the action of the isometric quantum channel $\mathcal{V}^{\prime}$ on the quantum systems $S_{2}^{}$ and $S_{4}^{\prime}$, while keeping the reference system $R^{\prime}$.
    \item Prepare the quantum systems $S_{0}^{\prime}$, $S_{2}^{\prime}$, $S_{3}^{\prime}$ and $S_{5}^{\prime}$ on quantum states $\tau_{0}$, $\tau_{2}$, $\tau_{3}$ and $\tau_{5}$, respectively.
    \item Discard the input local quantum systems at $t=1,3,4,6$.    
\end{enumerate}

Finally, the intervention for the computation of $\widetilde{I_c}(\Upsilon_1(\Lambda) ; \Upsilon_3 \circ \Upsilon_2)$, denoted as $\mathcal{I}_2^{\prime}$, is represented with the diagram in Fig.~\ref{fig:HOQMMI}(f) and defined as follows.
\begin{enumerate}[$(i)$]
    \item Repeat the first quantum operation in the definition of $\mathcal{I}_2$. That is, take the action of the fixed isometric channel $\mathcal{V}^{\prime}$ accordingly.
    \item Prepare the quantum systems $S_{2}^{\prime}$ and $S_{3}^{\prime}$ on states $\tau_{2}$ and $\tau_{3}$, respectively.
    \item Take the action of identity channels on the slots at $t=1$ and $t=5$.
\end{enumerate}

From the above recipe of quantum interventions it is possible to state the interventional formulation of the higher-order quantum Markov monogamy inequality as follows. 
\begin{corollary}[Interventional formulation of the higher-order quantum Markov monogamy inequality] Let $\Upsilon$ be a seven-steps higher-order Markovian quantum process. Let $\Omega_{1},\Omega_{1}^{\prime},\Omega_{2}$ and $\Omega_{2}^{\prime}$ be the quantum channels generated with the interventions $\mathcal{I}_{1},\mathcal{I}_{1}^{\prime}$,$\mathcal{I}_{2}$ and $\mathcal{I}_{2}^{\prime}$, respectively. Then, it holds that
\begin{align}\label{eq:interventional_HOQMMI}
&\left[H(S_{0}^{\prime}\rightarrow S_{6}^{})_{\Tr_R\circ\,\Omega_{1}^{\prime}}-H(S_{0}^{\prime}\rightarrow S_{6}^{}\otimes R)_{\Omega_{1}^{\prime}}\right]\nonumber\\+&\left[H(S_{1}^{\prime}\rightarrow S_{5}^{})_{\Tr_R\circ\,\Omega_{2}}-H(S_{1}^{\prime}\rightarrow S_{5}^{}\otimes R)_{\Omega_{2}}\right]\nonumber  
\\ 
&\geq \nonumber
\\
&\left[H(S_{0}^{\prime}\rightarrow S_{6}^{})_{\Tr_R\circ\,\Omega_{2}^{\prime}}-H(S_{0}^{\prime}\rightarrow S_{6}^{}\otimes R)_{\Omega_{2}^{\prime}}\right]\nonumber\\+&\left[H(S_{1}^{\prime}\rightarrow S_{5}^{})_{\Tr_R\circ\,\Omega_{1}}-H(S_{1}^{\prime}\rightarrow S_{5}^{}\otimes R)_{\Omega_{1}}\right] .
\end{align} 
\end{corollary}

\subsection{Comparison of higher-order quantum information inequalities}

Now that we have appropriately defined the higher-order quantum Markov monogamy inequality in its interventional form, we are also able to make some very simple comparison with the previous higher-order quantum information inequality in regard to the detection of non-Markovian effects.
Similarly to the model considered in the main text, we take under scrutiny a five-slots quantum comb consisting of unitary layers $\mathcal{U}_k$ as defined in Eq.~(\ref{unitary_matrix}), with $k=1,\dots,6$ (see Fig.~\ref{fig:HOQMMI}(a)). 

In turn, we consider a sort of higher-order version of marginalization for quantum processes defined by acting with the identity channel on the desired slots to be ignored.
In this way, one is able to transform the five-slots quantum comb in a three-slots quantum comb in a way that it would be possible to check the validity of the higher-order quantum information inequality for the reduced process. 

Importantly, there are two marginalization interventions for which higher-order Markovianity is preserved: $(a)$ Discarding the slots at $t=2$ and $t=4$. This intervention is referred here as $\mathcal{I}_a$. See Fig.~\ref{fig:HOQMMI}(g). $(b)$ Discarding the slots at $t=1$ and $t=5$. Similarly, This intervention is denoted by $\mathcal{I}_b$. See Fig.~\ref{fig:HOQMMI}(h). 

Thus, acting with the intervention scheme discussed in Sec.~\ref{witness} on the reduced process generated with $\mathcal{I}_a$ and $\mathcal{I}_b$, one is able to determine the gap in Eq.~(\ref{eq:thm_1}) for the given reduced processes. Such quantities are denoted by $\mathrm{DP}_a$ and $\mathrm{DP}_b$, respectively. In turn, the gap in higher-order quantum information  inequality is denoted by $\mathrm{HOQMMI}$. 

 \begin{figure}[t!] 
\centering   
\includegraphics[width=0.23\textwidth]{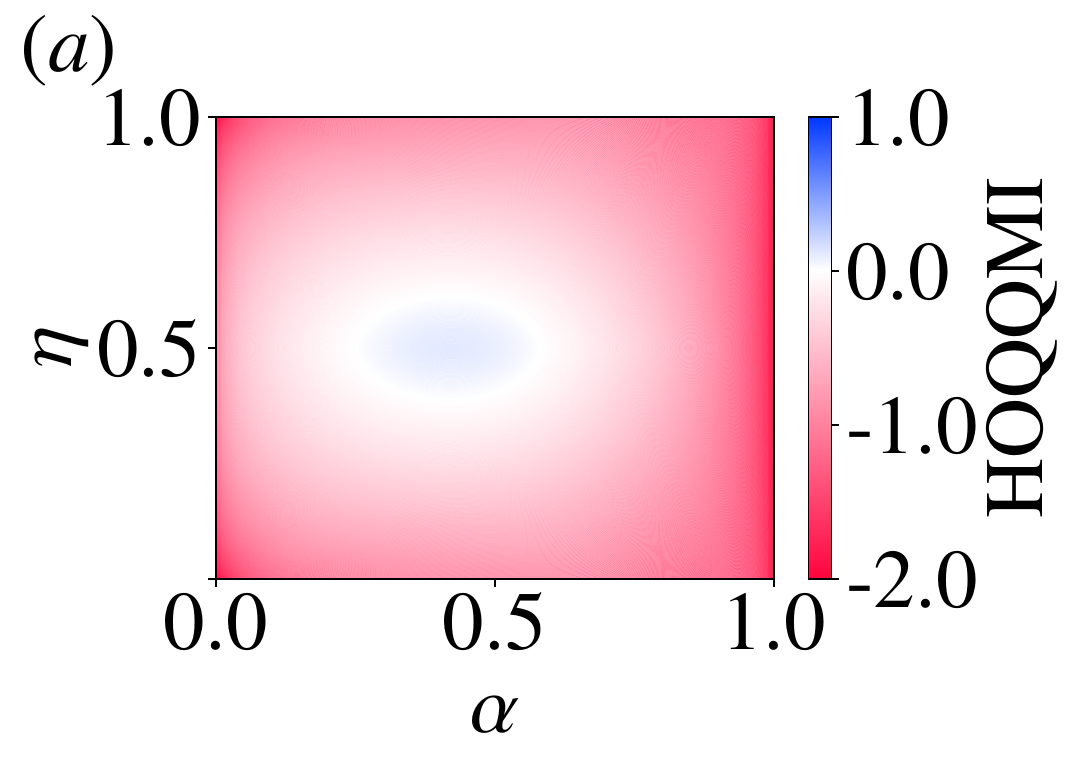} 
\includegraphics[width=0.23\textwidth]{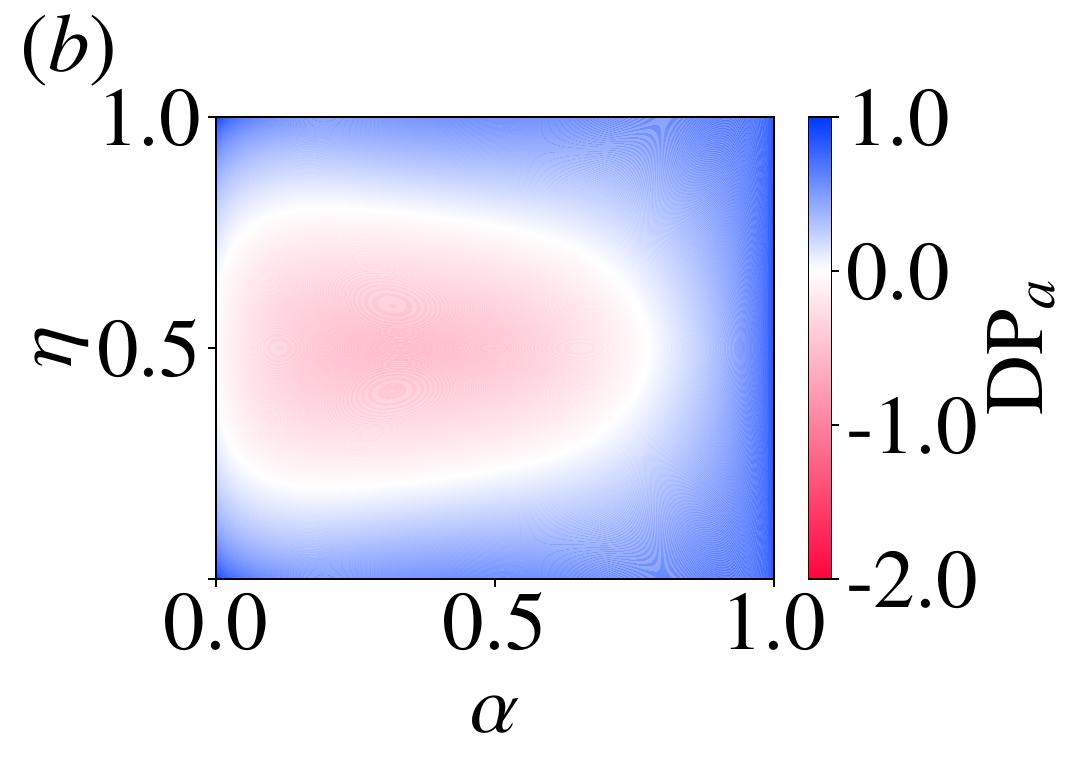} 
\includegraphics[width=0.23\textwidth]{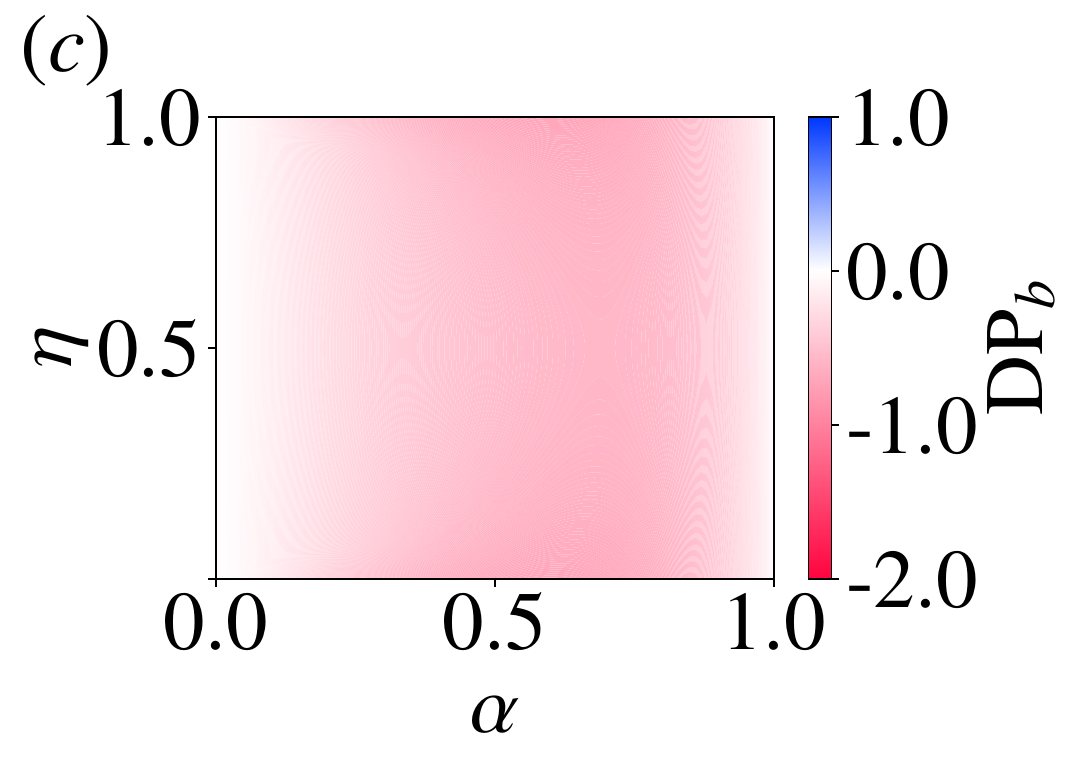} 
\caption{Violation of the higher-order quantum Markov monogamy inequality. Panel (a) shows the gap in the violation of Eq.~(\ref{eq:interventional_HOQMMI}). In turn, panels (b) and (c) show the violation of the marginal quantum processes, as discussed in the main text.}
    \label{fig:HOQMMI_plots}
\end{figure}

Figure~\ref{fig:HOQMMI_plots} shows the correspondent quantum information quantities depending upon the process parameter $\eta$ and the intervention parameter $\alpha$. Interestingly, the witnesses $\mathrm{HOQMMI}$ and $\mathrm{DP}_a$ have a complementary role for the model considered, in the sense that the region of parameters for which one is violated the other is valid and vice versa. See panels \ref{fig:HOQMMI_plots}(a) and \ref{fig:HOQMMI_plots}(b), respectively. In turn, the witness  $\mathrm{DP}_b$ provides a certification of higher-order non-Markovianity for almost all values of $\eta$ and $\alpha$, expect to the values parameters close to $\alpha=0$ and $\alpha=1$. 
 That is depicted in panel \ref{fig:HOQMMI_plots}(c). For those values, the witness $\mathrm{HOQMMI}$ provide a certification of the desired property, in a way that the three witness altogether perfectly witness the higher-order non-Markovianity for all values of the relevant parameters.


\bibliography{refs}
\end{document}